\documentclass[12pt]{article}

\advance \topmargin by -\headheight
\advance \topmargin by -\headsep
 
\evensidemargin \oddsidemargin
\usepackage{amsmath,amsthm,amssymb,bbm,bm}
\usepackage{graphicx,psfrag,epsf}
\usepackage{float,caption,subcaption}
\usepackage{multirow,threeparttable}
\usepackage{enumerate}
\usepackage{natbib}
\usepackage{hyperref}
\usepackage{url}
\usepackage{color}
\usepackage{comment}
\usepackage{kotex}

\usepackage{rotating}
\usepackage{titlesec}
\usepackage{appendix}

\newtheorem{theorem}{Theorem}

\def\bx{\mathbf{x}}
\def\by{\mathbf{y}}
\def\bX{\mathbf{X}}
\def\bY{\mathbf{Y}}
\def\bS{\mathbf{S}}
\def\bvarepsilon{\bm{\varepsilon}}
\def\bbeta{\bm{\beta}}
\def\beeta{\bm{\eta}}
\def\bSigma{\bm{\Sigma}}
\def\bP{\mathbf{P}}
\def\bQ{\mathbf{Q}}
\def\bG{\mathbf{G}}
\def\bGamma{\bm{\Gamma}}
\def\bOmega{\bm{\Omega}}

\def\tr{{{\rm tr}}}
\def\vec{{{\rm vec}}}
\def\Var{{{\rm Var}}}
\def\E{{{\rm E}}}

\newcommand{\appendixtitle}[3]{
  \begin{center}
    \Large \textbf{Supplementary Materials for ``Enhanced Response Envelope via Envelope Regularization"} \\[1em]
    \large \text{Oh-Ran Kwon and Hui Zou} \\[0.3em]
    \large \text{School of Statistics, University of Minnesota}
  \end{center}
  \vspace{3em} 
}
   
\newcommand{\blind}{1}

\begin{document}

\def\spacingset#1{\renewcommand{\baselinestretch}
{#1}\normalsize}

\spacingset{2}

\if1\blind 
{
  \title{\bf Enhanced Response Envelope \\ via Envelope Regularization}
  \author{Oh-Ran Kwon 
    and 
    Hui Zou
    \\
   School of Statistics, University of Minnesota}
  \date{}
  \maketitle
} \fi
 
\if0\blind
{
  \title{\bf Enhanced Response Envelope via Envelope Regularization}
  \author{author}
  \date{\today}
  \maketitle
} \fi

\bigskip
\begin{abstract}
    The response envelope model provides substantial efficiency gains over the standard multivariate linear regression by identifying the material part of the response to the model and by excluding the immaterial part. In this paper, we propose the enhanced response envelope by incorporating a novel envelope regularization term based on a nonconvex manifold formulation. It is shown that the enhanced response envelope can yield better prediction risk than the original envelope estimator. The enhanced response envelope naturally handles high-dimensional data for which the original response envelope is not serviceable without necessary remedies. In an asymptotic high-dimensional regime where the ratio of the number of predictors over the number of samples converges to a non-zero constant, we characterize the risk function and reveal an interesting double descent phenomenon for the envelope model. A simulation study confirms our main theoretical findings. Simulations and real data applications demonstrate that the enhanced response envelope does have significantly improved prediction performance over the original envelope method, especially when the number of predictors is close to or moderately larger than the number of samples.
    Proofs and additional simulation results are shown in the supplementary file to this paper.
\end{abstract}


\section{Introduction}
     The envelope model first introduced by \cite{cook2010envelope} is a modern approach to estimating an unknown regression coefficient matrix $\bbeta\in\mathbb R^{r\times p}$ in multivariate linear regression of the response vector $\by \in \mathbb R^r$ on the predictors $\bx\in\mathbb R^p$. It was shown by \cite{cook2010envelope} that the envelope estimator of $\bbeta$ results in substantial efficiency gains relative to the standard maximum likelihood estimator of $\bbeta$. The gains arise by identifying the part of the response vector that is material to the regression and by excluding the immaterial part in the estimation. The original envelope model has been later extended to the envelope model based on excluding immaterial parts of the predictors to the regression by \cite{cook2013envelopes}. 
    \cite{cook2013envelopes} then established the connection between the latter envelope model and partial least squares, providing a statistical understanding of partial least squares algorithms.

    The success of the envelope models and their theories motivated some authors to propose new envelope models by applying or extending the core idea of envelope modeling to various statistical models. 
    The two most common are the response envelope models and the predictor envelope models. The response envelope models (predictor envelope models) achieve estimation and prediction gains by eliminating the variability arising from the immaterial part of the responses (predictors) that is invariant to the changes in the predictors (responses). 
    Papers on response envelope models include the original envelope model \citep{cook2010envelope}, the partial envelope model \citep{su2011partial}, the scaled response envelope model  \citep{cook2013scaled}, the reduced-rank envelope model \citep{cook2015reducedrank}, the sparse envelope model \citep{su2016sparse}, the Bayesian envelope model \citep{khare2017bayesian}, the tensor response envelope model \citep{li2017parsimonious}, the envelope model for matrix variate regression \citep{ding2018}, and the spatial envelope model for spatially correlated data \citep{rekabdarkolaee2020new}. 
    Papers on predictor envelope models include the envelope model for predictor reduction \citep{cook2013envelopes}, the envelope model for generalized linear models and Cox's proportional hazard model \citep{cook2015foundations}, the scaled predictor envelope model \citep{cook2016scaled}, the envelope quantile regression model \citep{ding2020envelope}, the envelope model for the censored quantile regression  \citep{zhaoenvelopes}, 
    tensor envelope partial least squares regression \citep{zhang2017tensor}, and envelope-based sparse partial least squares regression  \citep{zhu2020envelope}. 
    Other than the response and predictor envelopes, \cite{cook2015simultaneous} developed the simultaneous envelope model for simultaneously reducing the response and the predictors. 
    For a comprehensive review of the envelope models, readers are referred to \cite{cook2018introduction}.

       High-dimensional data have become common in many fields. It is only natural to consider the performance of the envelope model under high dimensions. The likelihood-based method under both the response/predictor envelope model is not serviceable for high-dimensional data because the likelihood-based method requires the inversion of the sample covariance matrix of predictors. Hence, one has to find effective ways to mitigate this issue. 
    For the predictor envelope model, its connection to partial least squares provides one solution.
    Partial least squares \citep{de1993simpls} can be used for estimating the predictor envelope model \citep{cook2013envelopes}. The partial least squares algorithm is an iterative moment-based algorithm involving the sample covariance of predictors and the sample covariance between the response vector and predictors, which does not require inversion of the sample covariance matrix of predictors. In addition, the algorithm provides the root-$n$ consistent estimator of $\bbeta$ in the predictor envelope model with the number of predictors fixed \citep{chun2010sparse,cook2013envelopes} and can yield accurate prediction in the asymptotic high-dimensional regime when the response is univariate \citep{cook2019partial}. Motivated by this, \cite{zhu2020envelope} introduced envelope-based sparse partial least squares and showed the consistency of the estimator for the sparse predictor envelope model. \cite{zhang2017tensor} proposed a tensor envelope partial least squares algorithm, which provides the consistent estimator for the tensor predictor envelope model. 
    Another way to apply predictor envelope models for high-dimensional data is by selecting the principal components of predictors and then using likelihood-based estimation on the principal components. This simple remedy is adapted by \cite{rimal2019comparison} to compare the prediction performance of the likelihood-based predictor envelope method, principal component regression, and partial least squares regression for high-dimensional data. Their extensive numerical study showed that this simple remedy produced better prediction performance than principal component regression and partial least squares regression. The impact of high dimensions is more severe for the response envelope. There is far less work on making the response envelope model serviceable for high-dimensional data.  The Bayesian approach for the response envelope model \citep{khare2017bayesian} can handle high-dimensional data. The sparse envelope model \citep{su2016sparse} which performs variable selection on the responses can handle data with the sample size smaller than the number of responses, but still requires the number of predictors to be smaller than the number of sample size.

    In this paper, 
    we propose the enhanced response envelope for high-dimensional data by incorporating a novel envelope regularization term in its formulation. The envelope regularization term respects the fundamental idea of the original envelope model by considering the presence of the material and immaterial parts of the response in the model. The enhancements resulting from the enhanced response envelope are twofold. First, 
    our enhanced response envelope estimator can handle both low- and high-dimensional data, while the original envelope estimator can only handle low-dimensional data where the sample size $n$ is smaller than the number of predictors $p$. 
    From the connection between the original envelope estimator and the enhanced response envelope estimator in low-dimension, we extend the definition of the original envelope estimator to high-dimensional data by considering the limiting case of the enhanced response envelope estimator with a vanishing regularization parameter; see the discussion in Section \ref{sec:2-new-esti}.
    Second, we prove that the enhanced response envelope can reduce the prediction risk relative to the original envelope for all values of $n$ and $p$. Moreover, we study the asymptotics of the prediction risk for the original envelope estimator and the enhanced response envelope estimator when both $n,p\rightarrow \infty$ and their ratio converges to a nonzero constant $p/n\rightarrow \gamma\in (0,\infty)$. This kind of asymptotic regime has been considered in high-dimensional machine learning theory  \citep{el2018impact,dobriban2018high,liang2020,hastie2022surprises} for analyzing the behaviour of prediction risk of certain predictive models. We derive an interesting asymptotic prediction risk curve for the envelope estimator. The risk increases as $\gamma$ increases, and then decreases after $\gamma>1$. This phenomenon is known as double descent phenomenon in the machine learning literature \citep{belkin2019reconciling}.

    The enhanced response envelope is particularly valuable when dealing with data where the number of predictors is close to or moderately larger than the number of samples. In such scenarios, our asymptotic theoretical results and simulation findings demonstrate a substantial reduction in the prediction risk achieved by the enhanced response envelope compared with the original envelope method.

   The rest of the paper is organized as follows. We review the original envelope model and the corresponding envelope estimator in Section \ref{sec:2-review}. In Section \ref{sec:2-new-reg}, we introduce a new regularization term called the envelope regularization based on which we propose the enhanced response envelope in section \ref{sec:2-new-esti}. The enhanced response envelope estimator naturally provides a definition for the envelope estimator when $p>n$. Section \ref{sec:2-imple} describes how to implement this new method in practice. In Section \ref{sec:3-non-asymp}, we prove that the enhanced response envelope can yield better prediction risk than the original envelope for any $(n,p)$ pair. Considering $n,p\rightarrow \infty$ and $p/n\rightarrow \gamma\in (0,\infty)$, we derive the limiting prediction risk result of the original envelope and the enhanced response envelope in Section \ref{sec:3-double}. The theoretical results along with our simulation study in Section \ref{sec:sim} reveal the double descent phenomenon. 
    Real data analysis is presented in Section \ref{sec:real}. Proofs of theorems and supplementary tables are relegated to the supplementary file of this paper.

\section{Enhanced response envelope}\label{sec:2}
\subsection{Review of envelope model}\label{sec:2-review}

   \noindent \textit{\textbf{Envelope model }} Let us begin with the classical multivariate linear regression model of a response vector $\by \in \mathbb R^r$ given a predictor vector $\bx \in \mathbb R^p$: 
    \begin{equation}
    \begin{aligned} 
    \label{eq:linear-model}
    \by = \bbeta \bx + \bvarepsilon, ~\bvarepsilon \sim N(\mathbf 0,\bSigma),
    \end{aligned}
    \end{equation}
    where $\bvarepsilon$ is the error vector with a positive definite $\bSigma$ and independent to $\bx$. 
    An unknown matrix of regression coefficients is denoted as $\bbeta\in\mathbb{R}^{r\times p}$ and $\bx$ follows a distribution $P_\bx$ on $\mathbb R^p$ such that $\mathrm{E}(\bx)=\mathbf 0$ and $\mathrm{Cov}(\bx)=\bSigma_\bx$. We omit an intercept by assuming $\mathrm{E}(\by)=\mathbf 0$ for easy communication.

    The envelope model allows for the possibility that there is a part of the response vector that is unaffected by changes in the predictor vector. 
    Specifically, let $\mathcal{E}\subseteq\mathbb{R}^r$ be a subspace such that for all $\bx_1$ and $\bx_2$,
    \begin{equation} 
    \begin{aligned} 
    \label{eq:envlp-cond(1)}
    & && \text{(i)~}
    \bQ_{\mathcal{E}}\by| (\bx=\bx_1) \sim \bQ_{\mathcal{E}}\by| (\bx=\bx_2) ~\text{
    and }~\text{(ii)~}
    \bP_{\mathcal{E}}\by \perp \!\!\! \perp \bQ_{\mathcal{E}}\by|\bx,
    \end{aligned}
    \end{equation}
    where $\bP_{\mathcal{E}}$ is the projection onto $\mathcal{E}$ and $\bQ_{\mathcal{E}}=\mathbf I-\bP_{\mathcal{E}}$. Condition (i) states that the marginal distribution of $\bQ_{\mathcal{E}}\by$ is invariant to changes in $\bx$.  Condition (ii) says that $\bQ_{\mathcal{E}}\by$ does not affect $\bP_{\mathcal{E}}\by$ if $\bx$ is provided. 
    Conditions together imply that $\bP_{\mathcal{E}}\by$ includes the relevant dependency information of $\by$ on $\bx$ (the material part) while $\bQ_{\mathcal{E}}\by$ is the irrelevant information (the immaterial part).

    Let $\mathcal{B}=\text{span}(\bbeta)$. 
    The conditions in  \eqref{eq:envlp-cond(1)} hold if and only if 
    \begin{equation} 
    \begin{aligned} 
    \label{eq:envlp-cond(2)}
    \text{span}(\bbeta) = \mathcal{B}
    \subseteq \mathcal{E} ~\text{
    and }~ \bSigma = \bP_{\mathcal{E}} \bSigma
    \bP_{\mathcal{E}} + \bQ_{\mathcal{E}} \bSigma \bQ_{\mathcal{E}} .
    \end{aligned}
    \end{equation}
    The definition of an envelope introduced by \cite{cook2007dimension,cook2010envelope} formalizes the smallest subspace satisfying the conditions in \eqref{eq:envlp-cond(1)} using the equivalence relation of \eqref{eq:envlp-cond(1)} and \eqref{eq:envlp-cond(2)}. 
    The envelope is defined as the intersection of all subspaces $\mathcal{E}$ satisfying \eqref{eq:envlp-cond(2)} and is denoted by $\mathcal{E}_{\bSigma,\mathcal{B}}$, $\bSigma$-envelope of $\mathcal{B}$. 
    
    The envelope model arises by parameterizing the multivariate linear model in terms of the envelope $\mathcal{E}_{\bSigma,\mathcal{B}}$. The parameterization is as follows. Let $u = \text{dim}( \mathcal{E}_{\bSigma,\mathcal{B}} )$,
    $\bGamma \in \mathbb{R}^{r \times u}$ be any semi-orthogonal basis matrix for $\mathcal{E}_{\bSigma,\mathcal B}$, and $\bGamma_0\in\mathbb{R}^{r \times (r-u)}$ is any semi-orthogonal basis matrix for the orthogonal complement of $\mathcal{E}_{\bSigma,\mathcal B}$. 
    Then the multivariate linear model can be written as
    \begin{equation}
    \begin{aligned} 
    \label{eq:envlp-model}
    \by = \bGamma \beeta \bx + \bvarepsilon, ~\bvarepsilon \sim N(\mathbf 0, \bGamma \bOmega
    \bGamma^T + \bGamma_0 \bOmega_0 \bGamma_0^T),
    \end{aligned}
    \end{equation}
    where $\bbeta=\bGamma\beeta$ with $\beeta \in \mathbb{R}^{u\times p}$, and $\bOmega \in \mathbb{R}^{r\times r}$ and $\bOmega_0 \in
    \mathbb{R}^{(r-u)\times (r-u)}$ are symmetric positive definite matrices. Model \eqref{eq:envlp-model} is called the envelope model.  

    \medskip
    
    \noindent \textit{\textbf{Envelope estimator }} The parameters in the envelope model are estimated by maximizing the likelihood function  from model \eqref{eq:envlp-model}. Assume that $p{+}r < n$ and $u$ is the dimension of the envelope. Define $\bS_\bX=n^{-1} \bX^T\bX$, $\bS_\bY = n^{-1} \bY^T \bY$, $\bS_{\bY,\bX} = n^{-1}\bY^T\bX$, and $\bS_{\bY|\bX} = \bS_\bY - \bS_{\bY,\bX} \bS_\bX^{-1} \bS_{\bX,\bY}$, where $\bY\in\mathbb R^{n\times r}$ has rows $\by_i^T$ and $\bX\in\mathbb R^{n\times p}$ has rows $\bx_i^T$. 
     
    The envelope estimator of $\bbeta$ is determined as
    \begin{equation}\begin{aligned}\label{eq:std-env-est}
  &\hat{\mathcal E}_{\bSigma,\mathcal B} = \text{span} \{ \arg\min_{\bG\in \text{Gr}(r,u)} (\log |\bG^T \bS_{\bY|\bX}\bG| + \log |\bG^T \bS_\bY^{-1}\bG|)\}, 
    \end{aligned}\end{equation}
       where $\text{Gr}(r,u)=\{\bG\in\mathbb R^{r\times u}: \bG \text{ is a semi-orthogonal matrix}\}$. Define  $\hat\bGamma$ as any semi-orthogonal basis matrix for $\hat{\mathcal E}_{\bSigma,\mathcal B}$ and let $\hat\bGamma_0$ be any semi-orthogonal basis matrix for the orthogonal complement of $\hat{\mathcal E}_{\bSigma,\mathcal B}$. 
       The estimator of $\bbeta$ is given by
    \begin{equation}\begin{aligned}\label{eq:std-env-est2}
     & \hat\bbeta = \hat\bGamma\hat{\bGamma}^T \bS_{\bY,\bX}\bS_\bX^{-1},\\
     \end{aligned}\end{equation}
and $\bSigma$ is estimated by $\hat \bSigma=\hat \bGamma \hat \bOmega \hat \bGamma +\hat \bGamma_0^T \hat \bOmega_0 \hat \bGamma_0$ where
   \begin{equation}\begin{aligned}\label{eq:std-env-est3}
     & \hat\bOmega =  \hat{\bGamma}^T \bS_{\bY|\bX} \hat{\bGamma}, ~~
        \hat\bOmega_0 = \hat\bGamma_{0}^T \bS_\bY \hat\bGamma_{0}.
            \end{aligned}\end{equation}

\subsection{Envelope regularization}\label{sec:2-new-reg}
    In this section, we introduce the envelope regularization term that respects the fundamental idea of the envelope model by considering the envelope structure of the presence of material and immaterial parts, $\bP_{\mathcal{E}_{\bSigma,\mathcal B}} \by$ and $\bQ_{\mathcal{E}_{\bSigma,\mathcal B}} \by$, in the regression. 
    
    We define the envelope regularization term as
    \begin{equation}\label{eq:penalty}
        \rho(\beeta,\bOmega) = \text{tr}(\beeta^T \bOmega^{-1} \beeta).
    \end{equation}
    The envelope model distinguishes between $\bP_{\mathcal{E}_{\bSigma,\mathcal B}} \by$ and $\bQ_{\mathcal{E}_{\bSigma,\mathcal B}} \by$ in the estimation process because of the envelope structure \eqref{eq:envlp-cond(2)}. Specifically, the log-likelihood function of the envelope model is decomposed into two log-likelihood functions. One is the log-likelihood function for the multivariate regression of $\bGamma^T \by$ on $\bx$, $\bGamma^T \by = \beeta \bx + \bGamma^T\bvarepsilon$ where $\bGamma^T\bvarepsilon\sim N(\mathbf 0,\bOmega)$. The other is the log-likelihood function for the zero-mean model of $\bGamma_0^T \by$, $\bGamma_0^T \by = \bGamma_0^T\bvarepsilon$ where $\bGamma_0^T\bvarepsilon\sim N(\mathbf 0,\bOmega_0)$. The envelope regularization term \eqref{eq:penalty} is the function of $\beeta$ and $\bOmega$, the parameters in the likelihood for the material part of the envelope model. 
    The envelope regularization term \eqref{eq:penalty} 
    encourages shrinkage 
    on the coefficient after standardizing the material part of the regression 
    to have uncorrelated errors,  $\bOmega^{-1/2}\bGamma^T\by = \bOmega^{-1/2}\beeta \bx + \bOmega^{-1/2} \bGamma^T\bvarepsilon$ where $\bOmega^{-1/2} \bGamma^T\bvarepsilon\sim N(\mathbf 0, \mathbf I)$.

    The well-known existing regularizations are usually the function of $\bbeta$, such as ridge regularization. In contrast, the envelope regularization is the function of $\beeta$ and $\bOmega$ as it takes into account the relationship between $\beeta$ and $\bOmega$ from the envelope structure. 
    As a result, the envelope regularization is optimized over $\beeta$ and $\bOmega$ simultaneously, as shown in the next subsection.

    The appealing features of using the envelope regularization over non-regularization are that it improves prediction performance and is capable of handling high-dimensions. Moreover, when compared to usual regularizations that are solely a function of $\bbeta$, the envelope regularization offers both theoretical and numerical advantages. The envelope regularization allows the enhanced envelope estimator of $\bbeta$ and $\bSigma$ to be expressed in an analytic formula as shown in the next subsection, an advantage that does not present with the usual regularizations. This analytic expression not only facilitates theoretical analyses of the prediction risk, as discussed in Section \ref{sec:3}, but also ensures ease of implementation. The solution becomes readily solvable given an estimated envelope subspace, a process detailed in Subsection \ref{sec:2-imple}.

    \subsection{The proposed estimator}\label{sec:2-new-esti}
    We only assume that $r\leq n$ but $p$ is allowed to be bigger than $n$. The log-likelihood function under the envelope model \eqref{eq:envlp-model} is 
    \begin{equation*}\begin{aligned}
        \mathcal L_u(\beeta,\mathcal E_{\bSigma,\mathcal B},\bOmega,\bOmega_0) = 
        &-(nr/2) \log(2\pi) - (n/2)\log |\bGamma\bOmega\bGamma^T + \bGamma_0\bOmega_0\bGamma_0^T| \\
        &- (1/2)\sum_{i=1}^n (\by_i-\bGamma\beeta \bx_i)^T (\bGamma\bOmega\bGamma^T + \bGamma_0\bOmega_0\bGamma_0^T)^{-1} (\by_i-\bGamma\beeta \bx_i).
    \end{aligned}\end{equation*}
    By incorporating the envelope  regularization term $\rho$ given in the last subsection, 
    we propose the following 
    enhanced response envelope estimator via penalized maximum likelihood:
    \begin{equation}\begin{aligned}\label{eq:reg-mle}
        &\arg\max \{ \mathcal L_u(\beeta,\mathcal E_{\bSigma,\mathcal B},\bOmega,\bOmega_0) - \frac{n}{2}\lambda \cdot \rho( \beeta,\bOmega) \},
    \end{aligned}\end{equation}
    where $\lambda>0$ serves as a regularization parameter. 
    
    Let $\bS_\bX=n^{-1} \bX^T\bX$, $\bS_\bY = n^{-1} \bY^T \bY$, $\bS_{\bY,\bX} = n^{-1}\bY^T\bX$, $\bS^\lambda_\bX = \bS_\bX+\lambda \mathbf I$ and $\bS^\lambda_{\bY|\bX} = \bS_\bY - \bS_{\bY,\bX} (\bS_\bX^\lambda)^{-1} \bS_{\bX,\bY}$. After some basic calculations, \eqref{eq:reg-mle} can be expressed as
    \begin{equation}\begin{aligned}\label{eq:est-env}
    &\hat{\mathcal E}_{\bSigma,\mathcal B}(\lambda) = \text{span} \{ \arg\min_{\bG\in \text{Gr}(r,u)} (\log |\bG^T \bS^\lambda_{\bY|\bX}\bG| + \log |\bG^T \bS_\bY^{-1}\bG| ) \}, 
    \end{aligned}\end{equation} 
    where $\text{Gr}(r,u)=\{\bG\in\mathbb R^{r\times u}: \bG \text{ is a semi-orthogonal matrix}\}$. Let $\hat\bGamma_\lambda$ be any semi-orthogonal basis matrix for $\hat{\mathcal E}_{\bSigma,\mathcal B}(\lambda)$ and $\hat\bGamma_{0,\lambda}$ be any semi-orthogonal basis matrix for the orthogonal complement of $\hat{\mathcal E}_{\bSigma,\mathcal B}(\lambda)$. 
    The enhanced envelope estimator of $\bbeta$ is given by
 \begin{equation}\begin{aligned}\label{eq:est-env2}
             \hat\bbeta(\lambda) = \hat\bGamma_\lambda \hat{\bGamma}_\lambda^T \bS_{\bY,\bX}(\bS_\bX^\lambda)^{-1} 
    \end{aligned}\end{equation} 
 and $\bSigma$ is estimated by $\hat \bSigma(\lambda)=\hat \bGamma_{\lambda} \hat \bOmega(\lambda) \hat \bGamma_{\lambda} +\hat \bGamma_{0,\lambda}^T \hat \bOmega_0(\lambda) \hat \bGamma_{0,\lambda}$ where
   \begin{equation}\begin{aligned}\label{eq:est-env3}
     & \hat\bOmega(\lambda) = \hat{\bGamma}_\lambda^T \bS^\lambda_{\bY|\bX} \hat{\bGamma}_\lambda, ~~
        \hat\bOmega_0(\lambda) = \hat\bGamma_{0,\lambda}^T \bS_\bY \hat\bGamma_{0,\lambda}.
            \end{aligned}\end{equation}

    The enhanced response envelope estimator can naturally handle the case where $p \geq n{-}r$, while the original envelope estimator \eqref{eq:std-env-est} does not.  
    It is also well-known that regularization such ridge or lasso often destroys scale-equivanriance of the un-regularized models. Envelope regularization has the same effect. The enhanced envelope estimator is neither invariant nor equivariant under the rescaling of predictors. This contrasts with the original envelope estimator, which is scale equivariant.   

    Additionally, when the dimension of the envelope subspace is $r$ (i.e., $u=r$), the enhanced envelope estimator of $\bbeta$ reduces to the multivariate ridge estimator as $\hat{\mathcal E}_{\bSigma,\mathcal B}=\mathbb R^r$. A parallel relationship exists between the original envelope estimator and the standard multivariate estimator. Specifically, when $u=r$, the original envelope estimator of $\bbeta$ simplifies to the standard multivariate estimator.

    We consider taking the limit of the enhanced response envelope estimator with  $\lambda\rightarrow 0^+$, which can demonstrate the relationship between the original envelope estimator and the enhanced response envelope estimator: 
    \begin{equation}\begin{aligned}\label{eq:est-env-extstd}
    & \hat{\mathcal E}_{\bSigma,\mathcal B} = \text{span} \{ \arg\min_{\bG\in \text{Gr}(r,u)} (\lim_{\lambda\rightarrow 0^+} \log |\bG^T \bS^\lambda_{\bY|\bX}\bG| + \log |\bG^T \bS_\bY^{-1}\bG|)\} \\
        & \text{and }\hat\bbeta = \lim_{\lambda\rightarrow 0^+} \hat \bbeta(\lambda) .
        \end{aligned}\end{equation}
    When $p < n{-}r$, this definition recovers the original envelope estimator \eqref{eq:std-env-est}. This definition enables the use of the envelope estimator when $p \geq n{-}r$, without altering the definition of the original envelope estimator \eqref{eq:std-env-est} when $p < n{-}r$. Hence, we take \eqref{eq:est-env-extstd} as the extended definition of envelope estimator. In practice, we implement  \eqref{eq:est-env-extstd} by computing the enhanced response envelope estimator \eqref{eq:est-env} with a very small value of $\lambda$ such as $10^{-8}$. 
    
    As the enhanced response envelope estimator \eqref{eq:reg-mle} has flexibility on $\lambda$, the enhanced response envelope estimator with an appropriate choice of $\lambda$ can yield better prediction risk compared to the envelope estimator, which is discussed in Section \ref{sec:3}.

    \subsection{Implementation}\label{sec:2-imple}
    We discuss the Grassmannian manifold optimization required in \eqref{eq:est-env} in this subsection.
    Suppose that the dimension $u$ is specified and $\lambda$ is given. Our proposed estimator $\hat{\mathcal E}_{\bSigma,\mathcal B}(\lambda)$ for $\mathcal E_{\bSigma, \mathcal B}$ requires the optimization of non-convex objective over the Grassmannian $\mathcal G(u,r)$. Such a computation problem exists for the original envelope model as well. So far, the best known algorithm for solving envelope models is the algorithm introduced by \cite{cook2016note}. Thus, we employ their algorithm  to compute $\hat{\mathcal E}_{\bSigma,\mathcal B}(\lambda)$ in \eqref{eq:est-env}. Note that we standardize $\bX$ so that each column has a mean of 0 and a standard deviation of 1 before fitting any model. 
    We recommend standardizing the predictor $\bX$, but it is not always required. If predictors have different units or scales, standardization can help ensure that the envelope regularization applies uniformly to all predictors. However, if all predictor variables are on a similar scale, you may choose not to standardize them.
    
    In practice, the tuning parameter $\lambda$ and the dimension $u$ of the envelope are unknown. We use the cross-validation method to choose $(u,\lambda)$. 
    For the original envelope, $u$ can be selected by using AIC, BIC, LRT or cross-validation. BIC and LRT may be preferred for estimation as shown by simulations in \cite{su2013estimation} and cross-validation for prediction. Because the enhanced response envelope model has an additional tuning parameter $\lambda$, we propose to use cross-validation to find the best tuning parameter combination of $u$ and $\lambda$.

    We implement the enhanced response envelope method in the \texttt{eenvlp} package in R, which is available at \url{https://github.com/ohrankwon/eenvlp}.

\section{Theory}\label{sec:3}

    In this section, we show that the enhanced response envelope can reduce the prediction risk over the envelope for any $(n,p)$ pair. We then consider the asymptotic setting when $n,p\rightarrow\infty$ $p/n\rightarrow \gamma\in (0,\infty)$. This asymptotic regime has been considered in the literature \citep{el2018impact,dobriban2018high,liang2020,hastie2022surprises} for analyzing the behaviour of prediction risk of certain predictive models.
    
   In our discussion, we consider the case where $\mathcal E_{\bSigma, \mathcal B}$ is known, which has been assumed in the existing envelope papers to understand the core mechanism of envelope methodologies  \citep{cook2013envelopes,cook2015foundations,cook2015simultaneous}. 
     
\subsection{Reduction in prediction risk}\label{sec:3-non-asymp}
    Consider a test point $\bx_{\rm{new}} \sim P_\bx$. For an estimator $\hat\bbeta$, we define the prediction risk as 
    \begin{equation*}
    \begin{aligned}
    R(\hat\bbeta|\bX) & = \E[ \| \hat\bbeta \bx_{\rm new} - \bbeta \bx_{\rm new} \|^2 | \bX] \\
    & = \E[ \tr \{ (\hat\bbeta - \bbeta)\text{Cov}(\bx_{\rm new})(\hat\bbeta - \bbeta)^T \} | \bX]
    \end{aligned}
    \end{equation*}
    Note that this definition has the bias-variance decomposition, 
    \begin{equation*}
    \begin{aligned}
        R(\hat\bbeta|\bX) 
        & = \underbrace{\| \text{bias}(\vec(\hat\bbeta)|\bX) \|^2_{\bSigma_\bx \otimes \mathbf I_r}}_{\text{bias}^2} + \underbrace{\tr \{ \Var (\vec(\hat\bbeta) | \bX) {(\bSigma_\bx \otimes \mathbf I_r) }\}}_{\text{var}},
    \end{aligned}
    \end{equation*}
    where $\otimes$ is the Kronecker product of two matrices and $\| z \|^2_{\bSigma_\bx \otimes \mathbf I_r} = z^T (\bSigma_\bx \otimes \mathbf I_r) z$.

    Let $\bGamma$ be a semi-orthogonal basis matrix for $\mathcal E_{\bSigma,\mathcal B}$.
    Following the discussion in Section~\ref{sec:2-new-esti}, we take \eqref{eq:est-env-extstd} as the definition of the envelope estimator $\hat\bbeta_{\bGamma}$ .
    The prediction risk of $\hat\bbeta_{\bGamma}$ is
    \begin{equation*}
        \begin{aligned}
        R(\hat\bbeta_{\bGamma}|\bX) 
        & = \underbrace{\vec^T(\bbeta) [ \Pi_\bX \bSigma_\bx \Pi_\bX \otimes \mathbf I_r ] \vec(\bbeta)\phantom{\frac{a}{c}}}_{\text{bias}^2} + \underbrace{\frac{\tr(\bOmega)}{n} \tr( \bS_\bX^{+} \bSigma_\bx)}_{\text{var}},
        \end{aligned}
    \end{equation*}
    where $\Pi_\bX=\mathbf I_p-\bS_\bX^{+}\bS_\bX$ and $\bS_\bX^{+}$ represents the Moore-Penrose inverse of $\bS_\bX$.

    The prediction risk of the enhanced response envelope estimator $\hat\bbeta_{\bGamma} (\lambda)$ is
        \begin{equation}
        \begin{aligned}\label{eq:pred-risk_eenv}
          &R(\hat\bbeta_{\bGamma}(\lambda)|\bX) = E[ \| \hat\bbeta_{\bGamma}(\lambda) \bx_{\rm new} - \bbeta \bx_{\rm new} \|^2 | \bX] \\
        & = \underbrace{\lambda^2\vec^T(\bbeta) [  (\bS_\bX+\lambda \mathbf I)^{-1} \bSigma_\bx (\bS_\bX+\lambda \mathbf I)^{-1} \otimes \mathbf I_r ] \vec(\bbeta)\phantom{\frac{a}{c}}}_{\text{bias}^2} + \underbrace{\frac{\tr(\bOmega)}{n} \tr( \bSigma_\bx(\bS_\bX+\lambda \mathbf I)^{-1} \bS_\bX (\bS_\bX+\lambda \mathbf I)^{-1})}_{\text{var}}.
        \end{aligned}
    \end{equation}

    Theorem \ref{thm:exist} shows that using the envelope regularization always improves the prediction risk of the envelope model. 
    
    \begin{theorem}\label{thm:exist}
    There always exists a $\lambda>0$ such that $R( \hat{\bbeta}_{\bGamma}(\lambda) | \bX) < R ( \hat{\bbeta}_{\bGamma} | \bX)$.
    \end{theorem}

    The equation \eqref{eq:pred-risk_eenv} enables a direct comparison between the prediction risk of the enhanced envelope estimator $\hat\bbeta_{\bGamma} (\lambda)$ and that of the multivariate ridge estimator. As discussed in Section \ref{sec:2-new-esti}, fitting with $u=r$ reduces the enhanced envelope estimator to the ridge estimator. This indicates that the bias component in the prediction risk of the multivariate ridge estimator is identical to that of the enhanced envelope estimator. The variance component for the multivariate ridge is $\left(\tr(\bOmega)+\tr(\bOmega_0)\right) \cdot \tr( \bS_\bX^{+} \bSigma_\bx)/n$. Thus, as long as $\tr(\bOmega_0)>0$, the prediction risk of the enhanced response envelope estimator $\hat\bbeta_{\bGamma} (\lambda)$ is strictly lower compared to the multivariate ridge estimator. However, this advantage does not exist if the true envelope dimension is $r$ or $\tr(\bOmega_0)=0$.

\subsection{Limiting prediction risk and double descent phenomenon}\label{sec:3-double}

   The asymptotics of the envelope model are well-established in the case where $n$ diverges while $p$ is fixed \citep{cook2010envelope}, while not in a high-dimensional asymptotic setup. In this section, we examine the limiting risk of both the enhanced response envelope estimator and the envelope estimator in the high-dimensional asymptotic regime where $n,p\rightarrow \infty$ with $p/n \rightarrow \gamma \in (0,\infty)$. The number of response variables $r$ is fixed.  This kind of asymptotic regime has been considered in high-dimensional machine learning theory \citep{el2018impact,dobriban2018high,liang2020,hastie2022surprises} for analyzing the behaviour of prediction risk of certain predictive models.

    \begin{theorem}\label{thm:limit-envelop} Assume that $\bx$ has a bounded 4th moment and that $\bSigma_x=\mathbf I_p$ and $\tr(\beeta^T\beeta)=c^2$ for all $n,p$. Then as $n,p\rightarrow \infty$, such that $p/n\rightarrow \gamma\in(0,\infty)$, almost surely,
        $$
        R(\hat\bbeta_{\bGamma}|\bX) \rightarrow \begin{cases}
        \tr(\bOmega)\frac{\gamma}{1-\gamma} & \text{for}~\gamma <1 \\
            c^2(1-\frac{1}{\gamma}) + \tr(\bOmega)\frac{1}{\gamma-1} & \text{for}~\gamma>1,
            \end{cases}
        $$
    and
        $$
          R(\hat\bbeta_{\bGamma}(\lambda^\ast)|\bX) \rightarrow
            \tr(\bOmega)\gamma m(-\lambda^\ast), 
            ~~~~~~~~~~~~~~~~~~~~
        $$
    where $\lambda^\ast 
    =  \tr(\bOmega)\gamma/c^2$ and $m(z)=\frac{1-\gamma-z-\sqrt{(1-\gamma-z)^2-4\gamma z}}{(2\gamma z)}$.
    \end{theorem}      
    
    \begin{figure}[t!]
    \centering
    \includegraphics[width=0.8\linewidth]{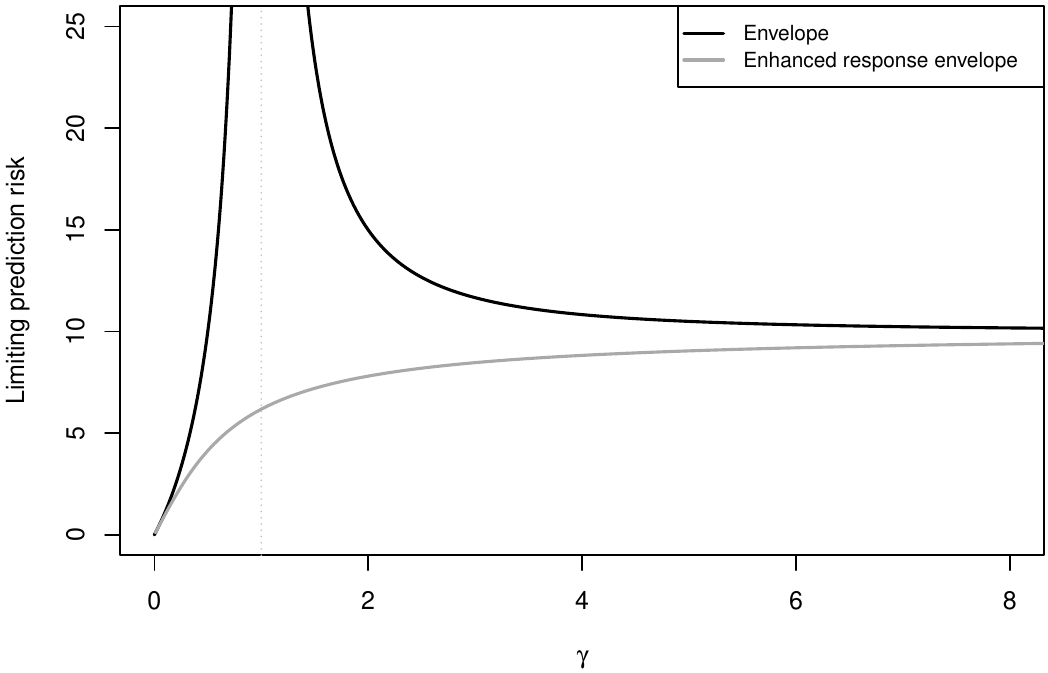}
    \caption{The limiting prediction risks of the enhanced response envelope with $\lambda^\ast = \tr(\bOmega)\gamma/c^2$ (gray solid line) and the envelope (black solid line), illustrating Theorem \ref{thm:limit-envelop}  when $\tr(\bOmega)=10$ and $\tr(\bbeta^T\bbeta)=10$.}
    \label{fig:limrisk-thm}
    \end{figure}
    
    Figure \ref{fig:limrisk-thm} visualizes the limiting prediction risk curves in Theorem \ref{thm:limit-envelop}. It plots the limiting risks of envelope (black solid line) and the enhanced response envelope with $\lambda^\ast=\tr(\bOmega)\gamma/c^2$ (dark-gray solid line), when  $\tr(\bOmega)=10$ and $\tr(\beeta^T\beeta)=10$.

    We have five remarks from Theorem \ref{thm:limit-envelop}. 
    The limiting risk of envelope increases before $\gamma=1$ and then decreases after $\gamma=1$. 
    The double descent phenomenon has been observed in popular methods such as neural networks, kernel machines and ridgeless regression \citep{belkin2019reconciling,hastie2022surprises}. It is interesting to see such a result is established in the envelope literature.   Second, the enhanced response envelope estimator always has a better asymptotic prediction risk than the envelope estimator (for any $c^2$, $\tr(\bOmega)$, and $\gamma$). Third, in Theorem \ref{thm:exist}, we show the existence of a $\lambda$ that gives a smaller prediction risk of the enhanced response envelope than the envelope estimator. In an asymptotic regime, we specify such a $\lambda$ value: $\lambda^\ast=\tr(\bOmega)\gamma/c^2$. Fourth, the gap between two limiting prediction risks, $\lim_{n,p\rightarrow \infty}R(\hat\bbeta_{\bGamma}|\bX)$ and ${\lim}_{n,p\rightarrow \infty}R(\hat\bbeta_{\bGamma}(\lambda^\ast)|\bX)$, 
    increases as $\gamma$ increases from $0$ to $1$. It is easy to see as $\frac{1}{1-\gamma} > m(-\lambda^\ast)$, $0<\gamma<1$. Lastly, as we can observe from Figure \ref{fig:limrisk-thm}, the gap between two limiting prediction risks is relatively larger when $\gamma$ is close to or moderately larger than $1$. The simulation results in the next section show similar trends to Figure \ref{fig:limrisk-thm} even with the estimated envelope subspace and correlated predictors.

    The uncorrelated covariates assumption allows us the obtain the explicit asymptotic risk function in Theorem \ref{thm:limit-envelop}. It is interesting to further relax this assumption by allowing a general dependence structure among covariates. In the next theorem, we relax  $\bSigma_\bx=\mathbf I_p$ to a general covariance matrix $\bSigma_\bx$. Motivated by \cite{dobriban2018high}, we assume that $\beeta \sim P_{\beeta}$ where $P_{\beeta}$ is the distribution on $\mathbb R^{u\times p}$ and $\E (\beeta^T\beeta)=c^2/p\cdot \mathbf I_p$. 
    
    \begin{theorem}\label{thm:limit-envelop-corr} Assume that $\bSigma_\bx^{-1/2}\bx$ has a bounded 12th moment and that $\max( \| \bSigma_\bx \|_2, \| \bSigma_\bx^{-1} \|_2) \leq M$ for all $n,p$. Assume that the spectral distribution of $\bSigma_\bx$ converges to a limiting distribution as $n,p\rightarrow \infty$. Then as $n,p\rightarrow \infty$, such that $p/n\rightarrow \gamma\in(0,\infty)$, almost surely,
        $$
        \mathrm E_{\beeta} \{ R(\hat\bbeta_{\bGamma}|\bX) \} \rightarrow \frac{c^2}{\gamma} \lim_{z\rightarrow 0^+} \frac{1}{v(-z)} + \tr(\bOmega) \lim_{z\rightarrow 0^+} \left(\frac{ v'(-z)}{ v(-z)^2} - 1\right)
        $$
    and
        $$
          \mathrm E_{\beeta} \{R(\hat\bbeta_{\bGamma}(\lambda^\ast)|\bX) \} \rightarrow
            \tr(\bOmega)\left( \frac{1}{\lambda^\ast v(-\lambda^\ast)}-1\right), 
            ~~~~~~~~~~~~~~~~~~~~
        $$
    where $\mathrm E_{\beeta}$ is the expectation over $\beeta$, $\lambda^\ast 
    =  \tr(\bOmega)\gamma/c^2$, $v$ is the Stieltjes transform of the limiting spectral distribution of $\bX \bX^T/n\in\mathbb R^{n\times n}$ and $v'$ is its derivative.
    \end{theorem}

    The above Theorem \ref{thm:limit-envelop-corr} is established under a general covariance structure for the predictors. However, there is no analytic form for $v$ or $v'$ in general. In order to better understand the theory, we conduct a simulation study to examine the asymptotic trends for correlated cases. Our simulation results suggest that the main trend shown in Figure \ref{fig:limrisk-thm} remains valid, although the actual form of the asymptotic risk function varies with the correlation structure.

\section{Simulation}\label{sec:sim}
    In this section, we use simulations to compare the performance of the enhanced response envelope estimator, the envelope estimator, partial least squares regression (PLSR), multivariate ridge regression, and the ad hoc envelope estimator. We mention the ad hoc envelope estimation because it has been used in the chemometrics literature for implementing an envelope model for high-dimensional data. The basic idea is to first use principal component analysis to reduce the dimension and then fit the standard envelop model on the principal components \citep{rimal2019comparison}. Additionally, we use simulations to have a numeric illustration of the double descent phenomenon to confirm the asymptotic theory.
    
    We consider a setting where $\by_i \in \mathbb R^3$ is generated from the model 
    $$
        \by_i = \bbeta \bx_i + \bvarepsilon_i, ~ \bvarepsilon_i \sim N(\mathbf 0,\bSigma),~i=1,\ldots,n,
    $$
    and $\bx_i\in\mathbb R^p$ is generated independently from $\bx_i\sim N(\mathbf 0,\bSigma_\bx(\rho))$ where  $(i,j)$th element of $\bSigma_\bx(\rho)\in\mathbb R^{p\times p}$ is $\rho^{|i-j|}$. The covariance matrix $\bSigma$ is set using three orthonormal vectors and has eigenvalues $10$, $8$, and $2$. 
    The columns of $\bGamma$ are the second and third eigenvectors of $\bSigma$. Each component of $\tilde{\beeta}\in\mathbb R^{2 \times p}$ is generated from the standard normal distribution. We then set $\beeta = \sqrt{10} \cdot \tilde{\beeta} / \| \tilde{\beeta} \|_F$.
    In this setting, $\text{dim}(\mathcal E_{\bSigma,\mathcal B})=2$, $\tr(\beeta^T\beeta) = 10$,  $\tr(\bOmega)=10$, and $\tr(\bOmega_0)=10$.

    \subsection{Prediction risk comparison}\label{sec:sim-2}
    In this simulation, we try three different cases of $n$ and $p$, that are $p<n$, $p\approx n$, and $p>n$. We compare the prediction risk of seven estimators: the enhanced response envelope estimator and the envelope estimator (each in two versions, depending on whether $\text{dim}(\mathcal E_{\bSigma,\mathcal B})$ is known or unknown), along with PLSR, the multivariate ridge regression, and the ad hoc envelope estimator. To numerically calculate the prediction risk, we employ the Monte Carlo simulations. As the prediction risk equals to $R(\hat\bbeta|\bX) = \E [ \E\{ \| \hat\bbeta \bx_{\rm new} - \bbeta \bx_{\rm new} \|^2 | \bX,\bY \} | \bX ]$, we compute it numerically using 
    $
        M^{-1} \sum_{m=1}^M \E[ \| \hat\bbeta \bx_{\rm new} - \bbeta \bx_{\rm new} \|^2 | \bX,\bY_m ],
    $
    where $i$-th row of $\bY_m \in \mathbb R^{n\times r}$ is generated from the conditional distribution of $\by$ given $i$-th row of $\bX\in\mathbb R^{n\times p}$.

    We use the R package \texttt{pls} to perform PLSR. For the ad hoc envelope estimator, we use the original envelope on the first few principal components explaining $97.5\%$ of the variations in the predictors and assume that $\text{dim}(\mathcal E_{\bSigma,\mathcal B})=2$ is known. If the number of selected principal components is equal to or bigger than $n-r$, we use the first $n-r-1$ principal components so that the original envelope estimator can be serviceable. 

    Given $(\bX,\bY_m)$, for the enhanced response envelope with a known $\text{dim}(\mathcal E_{\bSigma,\mathcal B})$ and for the multivariate ridge estimator, we perform ten-fold cross-validation on simulated data to select $\lambda$ among equally spaced 100 candidate $\lambda$-values in the scale of logarithm base 10. For the enhanced response envelope with an unknown $\text{dim}(\mathcal E_{\bSigma,\mathcal B})$, we again employ ten-fold cross-validation to choose the pair $(u,\lambda)$ from $u\in\{0,\ldots,r\}$ and the same set of 100 candidate $\lambda$-values. Also, ten-fold cross-validation is used to choose $u$ within $\{0,\ldots,r\}$ for the envelope estimator with an unknown $\text{dim}(\mathcal E_{\bSigma,\mathcal B})$ and to select the number of components for PLSR.

    We compute the envelope estimator for data with $n\leq p{+}r$ by taking a very small value of $\lambda=10^{-8}$ in the enhanced response envelope estimator; see the definition of the envelope estimator \eqref{eq:est-env-extstd} in Section \ref{sec:2-new-esti}. We then calculate the $\E[ \| \hat\bbeta \bx_{\rm new} - \bbeta \bx_{\rm new} \|^2 | \bX,\bY_m ]$ for each estimator. This process is repeated for each $m=1,\ldots,100$ ($M=100$). We then report the average of $\E[ \| \hat\bbeta \bx_{\rm new} - \bbeta \bx_{\rm new} \|^2 | \bX,\bY_m ]$ over $100$ replications, that is the prediction risk, with the standard error. Moreover, we report the bias and variance components of the prediction risk in Tables B.1-B.3, and the number of times each dimension was selected by the cross-validation among $100$ replications in Tables C.1-C.3, of the supplementary file.
    
    Tables \ref{tab:sim:less}, \ref{tab:sim:approx}, and \ref{tab:sim:greater} summarize the prediction risk of $p<n$, $p \approx n$, and $p>n$ cases, respectively. 
    We try different combinations of $n$, $p$, and $\rho$ with $n \in \{ 50,90,200 \}$ and $\rho\in \{0,0.8\}$. We choose $p$ so that $p/n\in \{ 0.1,0.3\}$ for $p<n$ cases, $p/n\in \{ 0.9,1.1\}$ for $p\approx n$, and $p/n\in \{ 3,10\}$ for $p>n$ cases.

    \begin{table}
    \scriptsize 
    \begin{center}
        \caption{\label{tab:sim:less} {\bf{(Cases of $\bm{p < n}$)}} Prediction risk. The standard error is given in parentheses.}
    \begin{tabular}{|c|c||ccccccc|} \hline
    $n$ &  $p$     & 
    \begin{tabular}[c]{@{}c@{}}Enhanced\\envelope\\(using true $u$)\end{tabular} & 
    \begin{tabular}[c]{@{}c@{}}Envelope\\(using true $u$)\end{tabular} & 
    \begin{tabular}[c]{@{}c@{}}Enhanced\\envelope\\(cv selected $u$)\end{tabular} & 
    \begin{tabular}[c]{@{}c@{}}Envelope\\(cv selected $u$)\end{tabular} & 
    PLSR & \begin{tabular}[c]{@{}c@{}}Ad hoc\\envelope\end{tabular} 
    & \begin{tabular}[c]{@{}c@{}}Ridge\\regression\end{tabular} \\  \hline\hline
    \multicolumn{9}{|l|}{{\bf Example 1.1}: ~$p/n=0.1$, $\rho=0$, $u=2$} \\ \hline\hline 
    50&5    & 1.56 (0.10) & 1.64 (0.10) & 1.67 (0.10) & 1.86 (0.12) & 2.45 (0.12) & 1.64 (0.10) & 1.97 (0.08) \\ \hline
    90&9    & 1.33 (0.05) & 1.46 (0.05) & 1.44 (0.07) & 1.60 (0.08) & 2.18 (0.08) & 1.46 (0.05) & 2.02 (0.06) \\ \hline
    200&20  & 1.15 (0.03) & 1.32 (0.03) & 1.21 (0.04) & 1.36 (0.05) & 1.72 (0.04) & 1.20 (0.04) & 1.86 (0.03) \\ \hline
    \multicolumn{9}{|l|}{{\bf Example 1.2}: ~$p/n=0.3$, $\rho=0$, $u=2$} \\ \hline\hline 
    50&15    & 4.01 (0.14) & 5.52 (0.22) & 4.20 (0.12) & 5.67 (0.18) & 5.86 (0.18) & 6.23 (0.19) & 4.91 (0.09)\\ \hline
    90&27    & 3.18 (0.08) & 5.27 (0.15) & 3.47 (0.10) & 5.61 (0.15) & 4.61 (0.12) & 4.62 (0.13) & 4.32 (0.06) \\ \hline
    200&60   & 3.06 (0.05) & 4.72 (0.09) & 3.13 (0.06) & 5.04 (0.10) & 4.42 (0.09) & 4.38 (0.06) & 4.45 (0.04) \\ \hline \hline
    \multicolumn{9}{|l|}{{\bf Example 1.3}: ~$p/n=0.1$, $\rho=0.8$, $u=2$} \\ \hline\hline 
    50&5    &  1.42 (0.07) & 1.68 (0.08) & 1.49 (0.06) & 2.08 (0.09) & 2.57 (0.06) & 1.68 (0.08) & 1.67 (0.06) \\ \hline
    90&9    &  1.14 (0.04) & 1.45 (0.05) & 1.29 (0.05) & 1.65 (0.09) & 2.34 (0.07) & 1.52 (0.05) & 1.57 (0.04)\\ \hline
    200&20  & 0.94 (0.03) & 1.34 (0.04) & 1.00 (0.04) & 1.50 (0.06) & 1.95 (0.04) & 1.29 (0.03) & 1.35 (0.03) \\ \hline
    \multicolumn{9}{|l|}{{\bf Example 1.4}: ~$p/n=0.3$, $\rho=0.8$, $u=2$} \\ \hline\hline 
    50&15    & 2.27 (0.07) & 5.54 (0.23) & 2.53 (0.08) & 4.58 (0.14) & 3.84 (0.14) & 3.29 (0.13) & 2.70 (0.06) \\ \hline
    90&27    & 2.27 (0.05) & 5.36 (0.15) & 2.53 (0.07) & 4.41 (0.11) & 3.92 (0.09) & 3.37 (0.08) & 2.89 (0.05) \\ \hline
    200&60  & 1.90 (0.03) & 4.75 (0.08) & 1.95 (0.04) & 5.11 (0.10) & 3.01 (0.06) & 2.96 (0.04) & 2.61 (0.03) \\ \hline
    \end{tabular}  
    \end{center}
    \end{table}

    \begin{table}
    \scriptsize 
    \begin{center}
        \caption{\label{tab:sim:approx} {\bf{(Cases of $\bm{p \approx n}$)}} Prediction risk. The standard error is given in parentheses.}
    \begin{tabular}{|c|c||ccccccc|} \hline
    $n$ &  $p$     & 
    \begin{tabular}[c]{@{}c@{}}Enhanced\\envelope\\(using true $u$)\end{tabular} & 
    \begin{tabular}[c]{@{}c@{}}Envelope\\(using true $u$)\end{tabular} & 
    \begin{tabular}[c]{@{}c@{}}Enhanced\\envelope\\(cv selected $u$)\end{tabular} & 
    \begin{tabular}[c]{@{}c@{}}Envelope\\(cv selected $u$)\end{tabular} & 
    PLSR & \begin{tabular}[c]{@{}c@{}}Ad hoc\\envelope\end{tabular}
    & \begin{tabular}[c]{@{}c@{}}Ridge\\regression\end{tabular} \\  \hline\hline
    \multicolumn{9}{|l|}{{\bf Example 2.1}: ~$p/n=0.9$, $\rho=0$, $u=2$} \\ \hline\hline 
    50&45    & 7.73 (0.12) & 161.84 (10.66) & 7.81 (0.11) & 10.18 (0.13) & 9.53 (0.16) & 13.60 (0.32) & 8.00 (0.07) \\ \hline
    90&81    & 6.98 (0.09) & 113.52 (4.06) & 7.13 (0.10) & 10.00 (0.00) & 9.45 (0.11) & 12.94 (0.22) & 7.70 (0.06) \\ \hline
    200&180   & 6.19 (0.05) & 115.23 (3.75) & 6.34 (0.06) & 10.00 (0.00) & 7.60 (0.10) & 11.89 (0.12) & 7.16 (0.03)\\ \hline
    \multicolumn{9}{|l|}{{\bf Example 2.2}: ~$p/n=1.1$, $\rho=0$, $u=2$} \\ \hline\hline 
    50&55    & 7.93 (0.13) & 62.55 (2.54) & 8.06 (0.12) & 11.36 (0.26) & 9.11 (0.12) & 14.10 (0.30) & 8.12 (0.07) \\ \hline
    90&99    & 7.34 (0.08) & 75.11 (2.36) & 7.52 (0.07) & 10.84 (0.22) & 9.40 (0.11) & 13.52 (0.23) & 7.94 (0.05) \\ \hline
    200&220   & 6.47 (0.05) & 94.71 (2.07) & 6.67 (0.06) & 10.29 (0.14) & 7.67 (0.10) & 12.16 (0.11) & 7.37 (0.03) \\ \hline
    \multicolumn{9}{|l|}{{\bf Example 2.3}: ~$p/n=0.9$, $\rho=0.8$, $u=2$} \\ \hline\hline 
    50&45    & 5.46 (0.15) & 162.13 (10.86) & 5.62 (0.14) & 14.14 (0.28) & 7.80 (0.16) & 8.73 (0.27) & 5.92 (0.11)\\ \hline
    90&81    & 3.93 (0.08) & 113.43 (4.18) & 4.08 (0.08) & 11.59 (0.00) & 6.98 (0.13) & 7.33 (0.14) & 4.90 (0.07) \\ \hline
    200&180  & 3.76 (0.04) & 116.53 (3.82) & 3.88 (0.05) & 9.95 (0.00) & 5.09 (0.06) & 7.04 (0.09) & 4.58 (0.03) \\ \hline
    \multicolumn{9}{|l|}{{\bf Example 2.4}: ~$p/n=1.1$, $\rho=0.8$, $u=2$} \\ \hline\hline 
    50&55    & 5.00 (0.14) & 67.01 (2.74) & 5.24 (0.14) & 14.67 (0.44) & 7.39 (0.22) & 8.47 (0.22) & 5.76 (0.11) \\ \hline
    90&99    & 4.08 (0.07) & 79.90 (2.62) & 4.44 (0.11) & 11.50 (0.27) & 6.16 (0.14) & 7.98 (0.16) & 4.97 (0.07) \\ \hline
    200&220   & 4.07 (0.05) & 99.06 (2.17) & 4.33 (0.06) & 11.28 (0.27) & 6.10 (0.06) & 7.77 (0.11) & 4.98 (0.03) \\ \hline
    \end{tabular}
    \end{center}
    \end{table}

    \begin{table} 
    \scriptsize
    \begin{center}
        \caption{\label{tab:sim:greater} {\bf{(Cases of $\bm{p > n}$)}} Prediction risk. The standard error is given in parentheses.}
    \begin{tabular}{|c|c||ccccccc|} \hline
        $n$ &  $p$     & 
    \begin{tabular}[c]{@{}c@{}}Enhanced\\envelope\\(using true $u$)\end{tabular} & 
    \begin{tabular}[c]{@{}c@{}}Envelope\\(using true $u$)\end{tabular} & 
    \begin{tabular}[c]{@{}c@{}}Enhanced\\envelope\\(cv selected $u$)\end{tabular} & 
    \begin{tabular}[c]{@{}c@{}}Envelope\\(cv selected $u$)\end{tabular} & 
    PLSR & \begin{tabular}[c]{@{}c@{}}Ad hoc\\envelope\end{tabular}
    & \begin{tabular}[c]{@{}c@{}}Ridge\\regression\end{tabular} \\  \hline\hline
    \multicolumn{9}{|l|}{{\bf Example 3.1}: ~$p/n=3$, $\rho=0$, $u=2$} \\ \hline\hline 
    50&150    & 9.77 (0.09) & 12.77 (0.12) & 9.88 (0.09) & 10.82 (0.15) & 10.62 (0.11) & 12.33 (0.14) & 9.56 (0.06) \\ \hline
    90&270    & 9.14 (0.04) & 12.48 (0.11) & 9.21 (0.04) & 9.99 (0.09) & 10.29 (0.07) & 12.39 (0.11) & 9.22 (0.03) \\ \hline
    200&600   & 8.64 (0.02) & 11.91 (0.07) & 8.73 (0.03) & 9.91 (0.09) & 10.07 (0.06) & 11.35 (0.08) & 8.97 (0.02) \\ \hline 
    \multicolumn{9}{|l|}{{\bf Example 3.2}: ~$p/n=10$, $\rho=0$, $u=2$} \\ \hline\hline 
    50&500    & 10.03 (0.03) & 10.49 (0.03) & 10.17 (0.05) & 10.37 (0.06) & 10.47 (0.05) & 10.44 (0.03) & 10.16 (0.05) \\ \hline
    90&900    & 9.85 (0.03) & 10.36 (0.03) & 9.99 (0.04) & 10.36 (0.05) & 10.41 (0.04) & 10.49 (0.03) & 9.96 (0.03) \\ \hline
    200&2000   & 9.72 (0.02) & 10.54 (0.05) & 9.87 (0.02) & 10.06 (0.02) & 10.12 (0.02) & 10.78 (0.04) & 9.77 (0.01) \\ \hline
    \multicolumn{9}{|l|}{{\bf Example 3.3}: ~$p/n=3$, $\rho=0.8$, $u=2$} \\ \hline\hline 
    50&150    & 7.84 (0.12) & 15.26 (0.38) & 7.96 (0.10) & 9.92 (0.22) & 9.52 (0.11) & 12.55 (0.26) & 7.71 (0.06) \\ \hline
    90&270    & 6.72 (0.10) & 14.56 (0.24) & 6.82 (0.10) & 8.22 (0.18) & 8.22 (0.21) & 11.67 (0.19) & 7.26 (0.04) \\ \hline
    200&600   & 6.03 (0.05) & 13.89 (0.17) & 6.15 (0.05) & 8.02 (0.07) & 7.62 (0.05) & 10.92 (0.14) & 6.88 (0.03) \\ \hline     
    \multicolumn{9}{|l|}{{\bf Example 3.4}: ~$p/n=10$, $\rho=0.8$, $u=2$} \\ \hline\hline 
    50&500    & 9.29 (0.07) & 11.65 (0.11) & 9.13 (0.10) & 9.78 (0.17) & 10.18 (0.14) & 11.79 (0.14) & 9.55 (0.06) \\ \hline
    90&900    & 9.16 (0.06) & 11.83 (0.10) & 9.33 (0.07) & 9.90 (0.15) & 10.63 (0.12) & 11.91 (0.12) & 9.56 (0.05) \\ \hline
    200&2000   & 8.62 (0.03) & 11.12 (0.05) & 8.74 (0.04) & 10.47 (0.06) & 10.31 (0.03) & 11.77 (0.10) & 9.30 (0.02) \\ \hline
    \end{tabular}
    \end{center}
    \end{table} 
            
    We have some comments on the simulation results for the prediction risk. 
    \begin{itemize}
 \item[(i)] The prediction risks of the enhanced response envelope with using the true $u$ (with cross-validated $u$) are consistently smaller than those of the envelope with using the true $u$ (with cross-validated $u$). Especially, we observe a substantial improvement in the prediction risks of the enhanced response envelope compared to the envelope when $n\approx p$. The improvement of the enhanced envelope over the envelope is more evident when the correlation among predictors is higher.
 
 \item[(ii)] When $u$ is selected by cross-validation, the enhanced response envelope model has higher prediction risk, which is expected due to the extra variability from the selection of $u$. The same trend for the envelope estimator with the cases of $p<n$. However, in cases where $p\approx n$ and $p>n$, the envelope estimator with cross-validated $u$ has lower prediction risk compared with the envelope estimator using the true $u$. This seems counter-intuitive. We examine the bias and variance decomposition of the prediction risk in  Tables B.1-B.3 and C.1-C.3 in the supplementary file. We observe that for the envelope estimator using the true $u$ results in lower bias but significantly higher variance when $p\approx n$ and $p>n$. Consequently, cross-validation chose a smaller envelope dimension to reduce the overall prediction risk of the envelope estimator, because the reduction in variance outweighs the increased bias. This empirical findings also confirm our earlier claim that the original envelope method is not equipped for handling high dimensional data. 
 
\item[(iii)] The prediction risks of the enhanced response envelope are consistently smaller than those of PLSR and the ad hoc envelope, except for the case of $p/n=0.1,~\rho=0$. 
 Comparing with the multivariate ridge regression, the enhanced response envelope (with cross-validated $u$) has smaller prediction risk, except for the cases of $p/n=10, ~\rho=0$ and $n=50, ~p/n=3$.
 
    \end{itemize}

    \subsection{Double descent confirmation}\label{sec:sim-double}

    In this subsection we use simulation to check the main messages in Theorems \ref{thm:limit-envelop} and \ref{thm:limit-envelop-corr}, i.e., the double descent phenomenon of the envelope model. Note that Theorems \ref{thm:limit-envelop} and \ref{thm:limit-envelop-corr} are established when considering that $\mathcal E_{\bSigma,\mathcal B}$ is known. In this simulation study, we do not use such information in the actual estimation in the simulation study. 
    
    We set $n\in \{ 200, 2000 \}$ and $\rho\in\{0,0.8\}$, and $p/n$ varies from $0.1$ to $5.0$. We compute the envelope and the enhanced response envelope with setting $u=2$ and $\lambda^\ast=\tr(\bOmega)p/(n c^2) = p/n$ on simulated data. We then calculate the prediction risk for each estimator. 
    Again, we fit $n \leq p{+}r$ data to the envelope estimator by taking a very small value of $\lambda=10^{-8}$ in the enhanced response envelope estimator.

\begin{figure}
    \centering
    \subfloat[\centering $n=2000,~\rho=0$\label{fig:limit-risk}]{{\includegraphics[width=8cm]{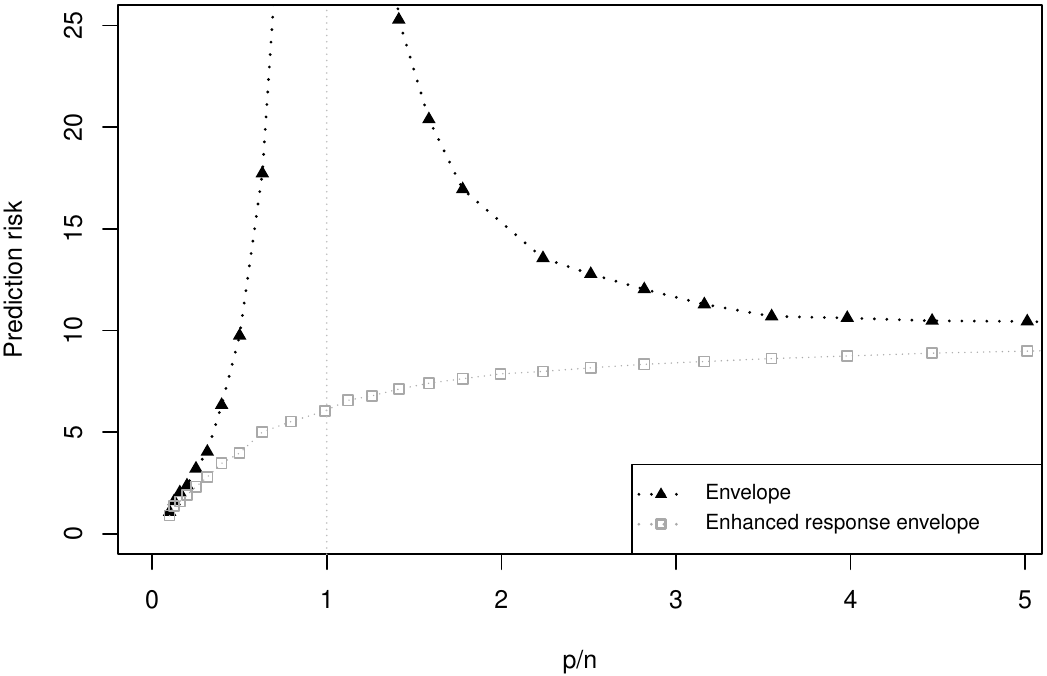} }}
    \subfloat[\centering $n=200,~\rho=0$\label{fig:limit-risk2}]{{\includegraphics[width=8cm]{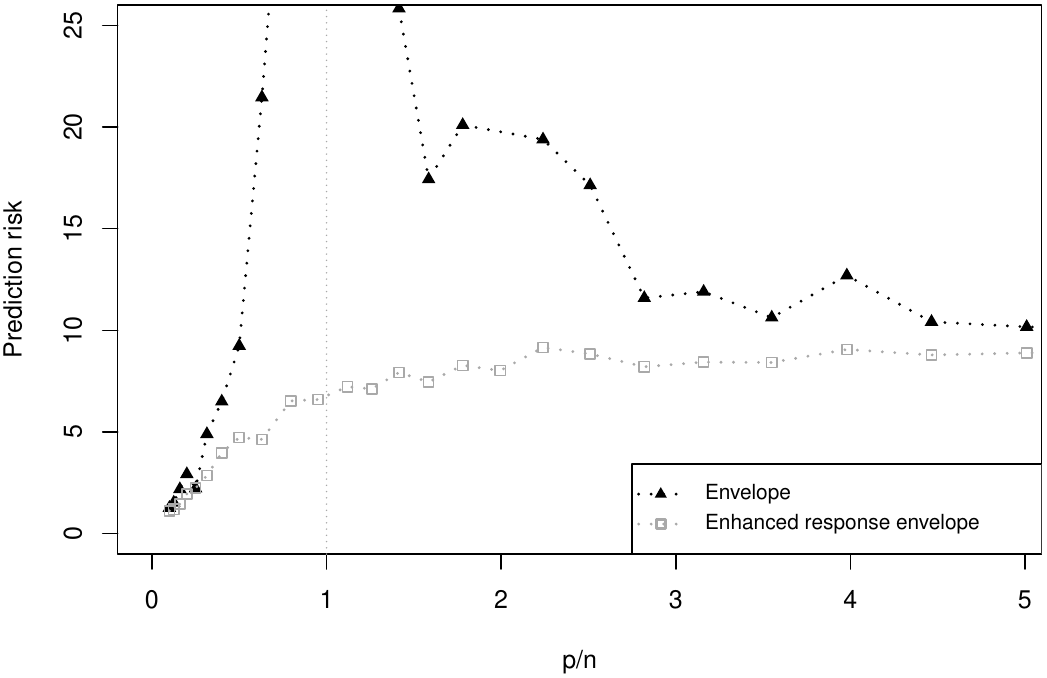} }}
    \\
    \subfloat[\centering $n=2000,~\rho=0.8$\label{fig:limit-risk_corr}]{{\includegraphics[width=8cm]{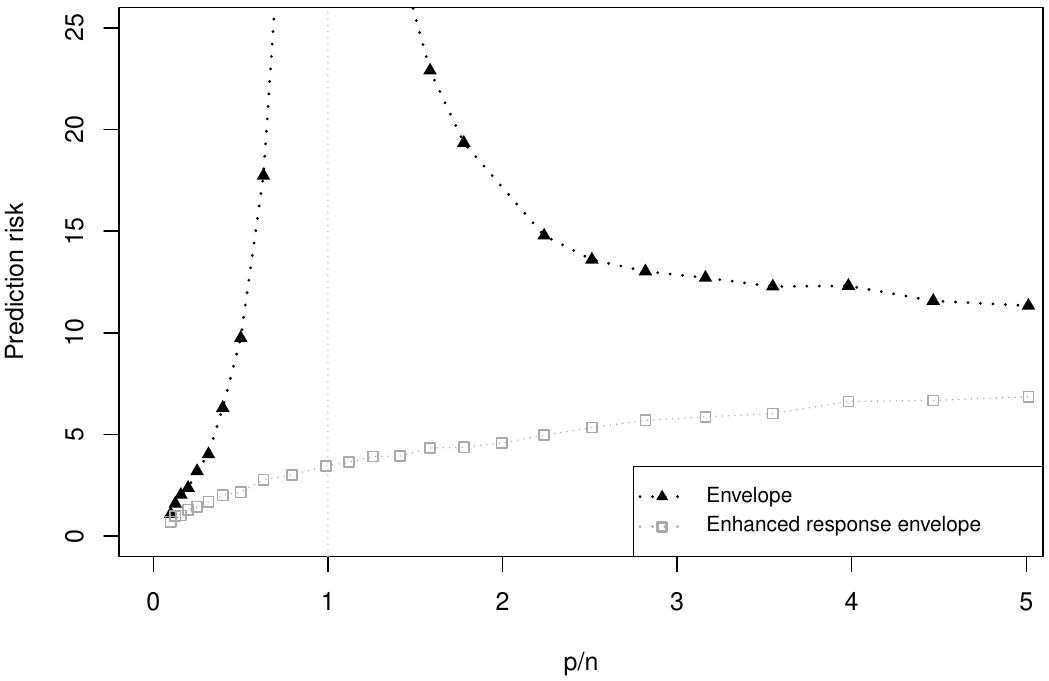} }}
    \subfloat[\centering $n=200,~\rho=0.8$\label{fig:limit-risk2_corr}]{{\includegraphics[width=8cm]{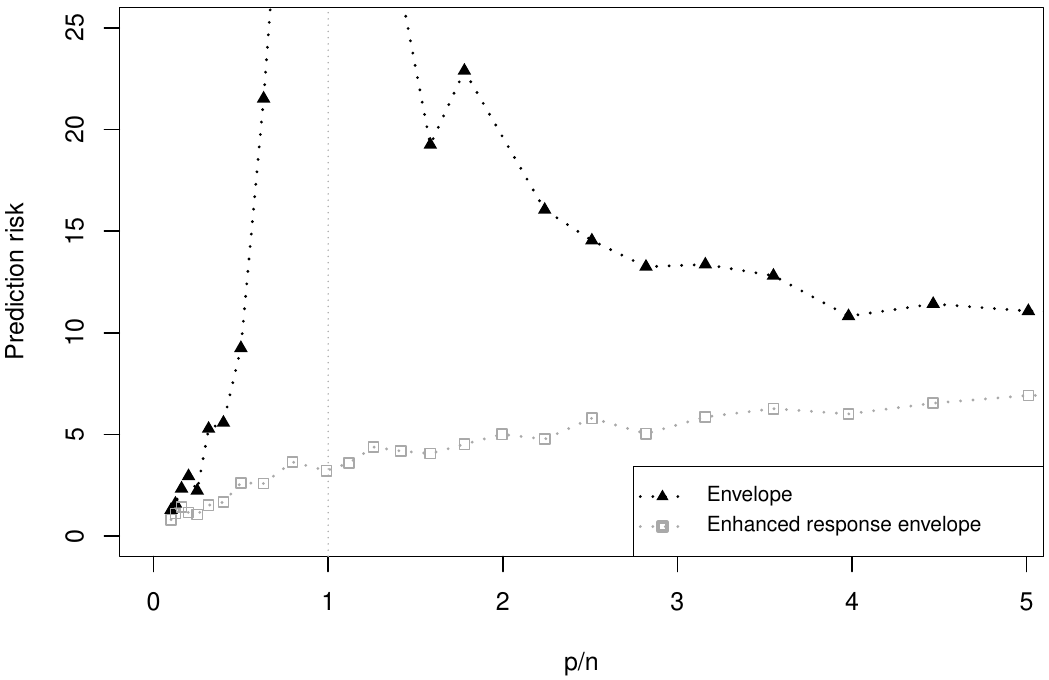} }}
    \caption{Prediction risk of the envelope and the enhanced response envelope with $\lambda^\ast=\tr(\bOmega)p/(n c^2)$ with various $p$ values. For $n\leq p{+}r$ data, we fit the envelope by taking a very small value of $\lambda=10^{-8}$ in the enhanced response envelope estimator.}
    \label{fig:leaky}
    \end{figure}

    Figures \ref{fig:limit-risk} and \ref{fig:limit-risk_corr} display the prediction risk curves of $n=2,000$ when $\rho=0$ and $\rho=0.8$, respectively. The gray rectangle points denote the prediction risk for the enhanced envelope estimator. The black triangle points are the prediction risk for the envelope estimator. We have four observations. 
    Firstly, we see that the enhanced response envelope 
   delivers a smaller prediction risk across the entire range of $p/n$ regardless of $n$. Secondly, the reduction of the enhanced response envelope estimator in terms of the prediction risk is much larger when $p\approx n$ compared to $p > n$ or $n > p$. Thirdly, the enhanced response envelope has better performance when predictors are correlated relative to uncorrelated predictors. Fourth, 
    we see a fascinating double descent prediction risk curve for the envelope model even with correlated predictors. 
    Figures \ref{fig:limit-risk2} and \ref{fig:limit-risk2_corr} plot the prediction risk curve when $n=200$. We see these figures exhibit the same messages for the much smaller sample size.

\section{A real data example}\label{sec:real}

 In Chemometrics, spectroscopy data are often used to characterize the chemical properties of matter, which is a rapid and cheap method compared to measuring the chemical contents using the conventional laboratory experiment method. PLSR is a very popular method in Chemometrics. Since its introduction, the envelope method has gained a lot of attention in the field \citep{cook2020}.  In this section, we use cereal's spectroscopy data to compare the prediction performance of the enhanced response envelope to the envelope, PLSR, and the ad hoc envelope estimator \citep{rimal2019comparison}.  The data are from \cite{varmuza2016introduction} available available in the \texttt{chemometrics} package in R (\url{https://cran.r-project.org/web/packages/chemometrics/index.html}). 
This data contain $15$ cereals $(n=15)$ on which $145$ infrared spectra were measured ($p=145$). We use the data to predict two chemical properties that are starch and ash ($r=2$).

As the sample size is small, we borrow the nested cross-validation idea \citep{wang2021honest,bates2021cross} to offer an honest comparison among competitor, in which an inner cross-validation is performed to tune a model and an outer cross-validation is performed to provide a prediction error of the tuned model. 
The observed prediction error calculated by this procedure, referred to as the honest LOOCV error, is an estimate of the generalization error which measures how well a tuned model predicts a randomly sample new test point. The honest LOOCV error is an almost unbiased estimator of the generalization error of the tuned model and performs better than other methods for estimating generalization errors \citep{wang2021honest}.

For $i \in \{1,2,\ldots,n\}$, we take the $i$th observation out from the data, and use the remaining $n{-}1$ observations to fit and tune models. We standardize $\bX \in \mathbb R^{(n-1)\times p}$ of the training set so that each column has a mean of 0 and a standard deviation of 1. On the $n-1$ data points, we perform the leave-one-out cross-validation to tune the model by selecting $(u,\lambda)$ from a fine grid of $u\in\{0,\ldots,r\}$ and $20$ equally spaced candidate $\lambda$-values in the scale of logarithm base 10 for the enhanced response envelope. For the envelope, we perform the leave-one-out cross-validation to choose $u$ from $\{0,\ldots,r\}$. 
For PLSR, the leave-one-out cross-validation is performed to select the number of components. 
For the ad hoc envelope, the first $q$ principal components explaining $99.5\%$ of the variation in the predictors are used as the predictors for simultaneous envelope estimation \citep{cook2015simultaneous}. The leave-one-out cross-validation is performed to select the tuning parameters on the simultaneous envelope. The dimension of $Y$-envelope is selected from $\{0,\ldots,r\}$ and the dimension of $X$-envelope is chosen from $\{0,\ldots,q\}$. 
  
The $i$th observation we take out at the beginning is set as the test set. We standardize $\bx_i$ of the test set using the mean and standard deviation of the training data. We then calculate the squared prediction error, $\| \by_i - \hat{\bbeta}^{(-i)} \bx_i \|_2^2/r$, where $\hat{\bbeta}^{(-i)}$ is the estimated regression coefficient derived from the training set. 
We repeat this process for $i=1,\ldots,n$ and report the honest LOOCV prediction error $\sum_{i=1}^n \| \by_i - \hat{\bbeta}^{(-i)} \bx_i \|_2^2/(nr)$ in Table \ref{tab:real}. We see that the enhanced response envelope estimator gives the smallest prediction errors among all competitors.

    \begin{table}[!t] 
    \begin{center}
        \caption{\label{tab:real} Prediction error for spectroscopy data. The frequencies of selected envelope dimension for the enhanced envelope estimator are: $\%(\hat u = 1) = 86.67\%$ and $\%(\hat u = 2) = 13.33\%$. The frequencies of selected envelope dimension for the envelope estimator are: $\%(\hat u = 1) = 93.33\%$ and $\%(\hat u = 2) = 6.67\%$.}
    \begin{tabular}{|ccccc|} \hline
    \begin{tabular}[c]{@{}c@{}}Enhanced\\envelope\end{tabular} & 
    Envelope &
    PLSR& \begin{tabular}[c]{@{}c@{}}Ad hoc\\envelope\end{tabular} &
    Ridge \\  \hline\hline
    0.298 & 0.405 & 0.395 & 0.381 & 0.340 \\ \hline
    \end{tabular} 
    \end{center}
    \end{table}

\section{Discussion}
    In this paper, we have developed a novel envelope regularization function which is used to define the enhanced envelope estimator. By asymptotic theory and extensive numerical examples, we have shown that the enhanced envelope estimator is indeed consistently better than the un-regularized envelope. The improvement is solely contributed by the novel envelope regularization term which involves two unknown model parameters: both $\beeta$ and $\bOmega$. In contrast, the usual regularization in regression is only a function of $\bbeta$. The asymptotic analysis of the risk function of the envelope reveals an interesting double descent phenomenon for the envelope model, which is also verified by simulations.    
         
    A direction for future research is to extend the methodology to the case when the number of responses $r$ $\rightarrow \infty$,  as this paper is focused on the case where $r$ is less than the number of samples and the number of predictors. \cite{su2016sparse} studied the response envelope for $r\rightarrow \infty$ but $p$ is fixed. When both $p, r$ diverge, there are additional technical issues to be addressed. For example, we may need another penalty term to handle the issues caused by the large $r$ in the model. This direction of research would be interesting to pursue.  

    Another direction for further research is to relax the normal error assumption which is a part of the standard envelope model. The present paper uses the same setup. \citet{su2012inner} demonstrated that the original envelope estimator attains asymptotic normality and $\sqrt{n}$-consistency under non-normal errors, provided certain conditions are met, as $n$ increases and $p$ remains fixed. Recently, \citet{HuberEnv} derived the enveloped Huber regression as a method for improving the standard envelope regression when the errors are non-normal and even potentially heteroscedastic. In a future paper, we will investigate the use of envelope regularization in the enveloped Huber regression context.

\if1\blind
{
\section*{Ackowledgements}
We would like to express our sincere gratitude to the Associate Editor and reviewers for their insightful comments and suggestions that significantly improved the quality of this manuscript. We also thank the Reproducibility Editor for their suggestions, which have improved the reproducibility of our code. Additionally, we thank R. Dennis Cook and Tate Jacobson for helpful discussions and feedback. Zou's research is supported in part by NSF grant 2220286.

\section*{Disclosure statement} 
The authors report there are no competing interests to declare. 
} \fi


\clearpage
\appendix
\begin{appendices}

\appendixtitle{Appendix}

~In this supplementary file we present the proofs of main theorems in the paper as well as some additional simulation results.

\section{Proofs of Theorems}\label{app:A}

\subsection{Proof of Theorem 1}
 
    Note that
        \begin{equation*}
        \begin{aligned}
            &R( \hat{\bbeta}_{\bGamma}(\lambda) | \bX)  = 
            \lambda^2 \tr(\bbeta  (\bS_\bX+\lambda \mathbf I)^{-1} \bSigma_\bx (\bS_\bX+\lambda \mathbf I)^{-1} \bbeta^T) + \frac{\tr(\bOmega)}{n} \tr( \bSigma_\bx (\bS_\bX+\lambda \mathbf I)^{-1}\bS_\bX (\bS_\bX+\lambda \mathbf I)^{-1}).
        \end{aligned}
        \end{equation*}
        Let $\mathbf M_\lambda = (\bS_\bX+\lambda \mathbf I)^{-1}\bSigma_\bX (\bS_\bX+\lambda \mathbf I)^{-1}$ and $\mathbf N_\lambda = \bS_\bX (\bS_\bX+\lambda \mathbf I)^{-1}\mathbf M_\lambda + \mathbf M_\lambda (\bS_\bX+\lambda \mathbf I)^{-1}\bS_\bX$. 
        
        We have
        \begin{equation*}
        \begin{aligned}
            &\frac{\partial}{\partial \lambda }R( \hat{\bbeta}_{\bGamma}(\lambda) | \bX)  \\
            = ~ & 
            2\lambda \tr(\bbeta^T\bbeta \mathbf M_\lambda)
            - \lambda^2 \tr(\bbeta^T\bbeta (\bS_\bX+\lambda \mathbf I)^{-1} \mathbf M_\lambda)
            - \lambda^2 \tr(\bbeta^T\bbeta  \mathbf M_\lambda (\bS_\bX+\lambda \mathbf I)^{-1}) \\
            & - \frac{\tr(\bOmega)}{n} \tr(\bS_\bx (\bS_\bX+\lambda \mathbf I)^{-1} \mathbf M_\lambda)
            - \frac{\tr(\bOmega)}{n} \tr(\bS_\bx \mathbf M_\lambda (\bS_\bX+\lambda \mathbf I)^{-1})
            \\
            = ~ & 
            \lambda \tr(\bbeta^T\bbeta (\bS_\bX+\lambda \mathbf I) (\bS_\bX+\lambda \mathbf I)^{-1}\mathbf M_\lambda) + \lambda \tr(\bbeta^T\bbeta \mathbf M_\lambda (\bS_\bX+\lambda \mathbf I)^{-1} (\bS_\bX+\lambda \mathbf I)) \\
            & - \lambda^2 \tr(\bbeta^T\bbeta (\bS_\bX+\lambda \mathbf I)^{-1} \mathbf M_\lambda)
            - \lambda^2 \tr(\bbeta^T\bbeta  \mathbf M_\lambda (\bS_\bX+\lambda \mathbf I)^{-1}) \\
            & - \frac{\tr(\bOmega)}{n} \tr(\bS_\bx (\bS_\bX+\lambda \mathbf I)^{-1} \mathbf M_\lambda)
            - \frac{\tr(\bOmega)}{n} \tr(\bS_\bx \mathbf M_\lambda (\bS_\bX+\lambda \mathbf I)^{-1})
            \\
            = ~ & 
            \lambda \tr(\bbeta^T\bbeta \bS_\bX (\bS_\bX+\lambda \mathbf I)^{-1}\mathbf M_\lambda) + \lambda \tr( \bbeta^T\bbeta \mathbf M_\lambda (\bS_\bX+\lambda \mathbf I)^{-1} \bS_\bX) \\
            & - \frac{\tr(\bOmega)}{n} \tr(\bS_\bx (\bS_\bX+\lambda \mathbf I)^{-1} \mathbf M_\lambda)
            - \frac{\tr(\bOmega)}{n} \tr(\bS_\bx \mathbf M_\lambda (\bS_\bX+\lambda \mathbf I)^{-1}) 
            \\
            = ~ & 
            \tr \left( (\lambda\bbeta^T\bbeta - \tr(\bOmega)/n \cdot \mathbf I ) \bS_\bX (\bS_\bX+\lambda \mathbf I)^{-1}\mathbf M_\lambda\right) +  \tr\left( ( \lambda \bbeta^T\bbeta- \tr(\bOmega)/n \cdot \mathbf I ) \mathbf M_\lambda (\bS_\bX+\lambda \mathbf I)^{-1}\bS_\bX \right)             \\
            = ~ & 
            \tr \left( (\lambda\bbeta^T\bbeta - \tr(\bOmega)/n \cdot \mathbf I ) ( \bS_\bX (\bS_\bX+\lambda \mathbf I)^{-1}\mathbf M_\lambda + \mathbf M_\lambda (\bS_\bX+\lambda \mathbf I)^{-1}\bS_\bX ) \right) 
            \\
            =~ & \tr \left( (\lambda\bbeta^T\bbeta - \tr(\bOmega)/n \cdot \mathbf I ) \mathbf N_\lambda \right)
            \\
             \leq ~ & \sum_{i=1}^p
            \left( \lambda \cdot \sigma_i(\bbeta^T \bbeta) - \frac{\tr(\bOmega)}{n} \right) \sigma_i(\mathbf N_\lambda ),
        \end{aligned}
        \end{equation*}
        where $\sigma_i(\mathbf M)$ denotes the $i$-th largest eigenvalue of $\mathbf M$. The last inequality above comes from the generalized Ruhe’s trace inequality. 

        Since $\mathbf N_\lambda$ is a positive semi definite matrix, then $\frac{\partial}{\partial \lambda }R( \hat{\bbeta}_{\bGamma}(\lambda) | \bX) <0$ if $\lambda < \tr(\bOmega) / (n  \sigma_1\left( \bbeta^T \bbeta) \right)$, and consequently, $R( \hat{\bbeta}_{\bGamma}(\lambda) | \bX) $ is a monotonically decreasing function if $0\leq \lambda \leq \tr(\bOmega) / (n  \sigma_1\left( \bbeta^T \bbeta) \right)$. 
        Therefore, we have
        \begin{equation*}
        \begin{aligned}
            &R( \hat{\bbeta}_{\bGamma}(\lambda) | \bX)  < 
            \frac{\tr(\bOmega)}{n} \tr( \bSigma_\bx  \bS_\bX^{+}), 
        \end{aligned}
        \end{equation*}
        when $0< \lambda < \tr(\bOmega) / (n  \sigma_1\left( \bbeta^T \bbeta) \right)$. Since 
        $$
            \frac{\tr(\bOmega)}{n} \tr( \bSigma_\bx  \bS_\bX^{+}) \leq R( \hat{\bbeta}_{\bGamma} | \bX),
        $$
        we prove the theorem.

\subsection{Proof of Theorem 2}

    As $\bSigma_\bx = \mathbf I$,
        \begin{equation*}
        \begin{aligned}
        &R(\hat\bbeta_{\bGamma}|\bX) 
         = \vec^T(\bbeta) [ \Pi_\bX \otimes \mathbf I_r ] \vec(\bbeta) + \frac{\tr(\bOmega)}{n} \tr( \bS_\bX^{+} ), \\
         &R( \hat{\bbeta}_{\bGamma}(\lambda) | \bX)  = 
            \lambda^2 \tr(\bbeta  (\bS_\bX+\lambda \mathbf I)^{-2}\bbeta^T ) + \frac{\tr(\bOmega)}{n} \tr(  \bS_\bX (\bS_\bX+\lambda \mathbf I)^{-2}),
        \end{aligned}
    \end{equation*}
        where $\Pi_\bX=\mathbf I_p-\bS_\bX^{+}\bS_\bX$. 
Our proofs below follow that of \cite{hastie2022surprises}.  
        
        \subsubsection{Proof for envelope estimator when \texorpdfstring{$\gamma<1$}{gamma<1}}
        Let us consider the case where  $p/n\rightarrow \gamma \in (0,1)$. 
        From Theorem 1 of \cite{bai2008limit}, $\sigma_{\min} (\bS_\bX) \geq (1-\sqrt{\gamma})^2/2$ and $\sigma_{\max} (\bS_\bX) \leq 2(1+\sqrt{\gamma})^2$ almost surely for all sufficiently large $n$. 
        Therefore, in this case, $\bS_\bX$ is invertible and the bias term of $R(\hat\bbeta_{\bGamma}|\bX)$ is 0,  almost surely. 
        
        The variance term of $R(\hat\bbeta_{\bGamma}|\bX)$ is 
        $$
            \frac{\tr(\bOmega)}{n} \tr( \bS_\bX^{+} ) = \frac{p\cdot\tr(\bOmega)}{n} \int \frac{1}{s} d F_{\bS_\bX}(s),
        $$
        where $F_{\bS_\bX}(s)$ is the spectral measure of $\bS_\bX$. By the Marchenko-Pastur theorem, which says that $F_{\bS_\bX} \rightarrow F_\gamma$, and the Portmanteau theorem, 
        $$
          \int_{(1-\sqrt{\gamma})^2/2}^{2(1+\sqrt{\gamma})^2/} \frac{1}{s} d F_{\bS_\bX}(s)
         \rightarrow 
          \int_{(1-\sqrt{\gamma})^2/2}^{2(1+\sqrt{\gamma})^2/} \frac{1}{s} d F_{\gamma}(s) = \int \frac{1}{s} d F_{\gamma}(s).
        $$
        The equality is because the support of $F_\gamma$ is $[(1-\sqrt{\gamma})^2,(1+\sqrt{\gamma})^2]$. We can also remove the upper and lower limits of integration on the left-hand side by Theorem 1 of \cite{bai2008limit}. Thus, combining above results, we arrive at 
        $$
            R(\hat\bbeta_{\bGamma}|\bX) \rightarrow \gamma \cdot \tr(\bOmega) \int  \frac{1}{s} d F_{\gamma}(s).
        $$
        
    The Stieltjes transformation of $F_\gamma$ is given by 
    $$
    m(z) = \int \frac{1}{s-z} d F_\gamma (s) = \frac{(1-\gamma-z) - \sqrt{(1-\gamma-z)^2 - 4\gamma z})}{2\gamma z},
    $$
    for any real $z<0$. By taking the limit $z\rightarrow 0^{-}$, the proof is completed. 
    \subsubsection{Proof for envelope estimator when \texorpdfstring{$\gamma>1$}{gamma>1}}
    
    The variance term of $R(\hat\bbeta_{\bGamma}|\bX)$ is 
    $$
            \frac{\tr(\bOmega)}{n} \tr( \bS_\bX^{+} ) = \frac{\tr(\bOmega)}{n} \tr( (\bX \bX^T/n)^{+} ) = \frac{\tr(\bOmega)}{p} \tr( (\bX \bX^T/p)^{+} ).
    $$
    Considering $n/p\rightarrow \tau = 1/
    \gamma <1$, by the same arguments from the proof above, we conclude that
    $$
        \frac{\tr(\bOmega)}{n} \tr( \bS_\bX^{+} ) \rightarrow \tr(\bOmega) \frac{1}{\gamma-1}. 
    $$
    
    Let $\bbeta = [\mathbf b_1^T ~ \ldots ~ \mathbf b_r^T]$. The bias term is
    $$
    \vec^T(\bbeta) [ \Pi_\bX \otimes \mathbf I_r ] \vec(\bbeta) = \sum_{i=1}^r \mathbf b_i^T \Pi_\bX \mathbf b_i = \sum_{i=1}^r \lim_{z\rightarrow 0^+} z \mathbf b_i^T (\bS_\bX + z\mathbf I)^{-1} \mathbf b_i.
    $$
    
    From Theorem 1 of \cite{bai2007asymptotics}, we have that 
    $$
        z \mathbf b_i^T (\bS_\bX + z \mathbf I)^{-1} \mathbf b_i \rightarrow z \int \frac{1}{s + z} F_\gamma(s) = z \|\mathbf b_i\|^2 m(-z) ~ \text{  a.s.},
    $$
    for any $i=1,\ldots,r$. We further have that
    $$
        \sum_{i=1}^r z \mathbf b_i^T (\bS_\bX + z \mathbf I)^{-1} \mathbf b_i \rightarrow z c^2 m(-z) ~ \text{  a.s}.
    $$
    
    By the Arzela-Ascoli  theorem and the Moore-Osgood theorem, we exchange limits and arrive at
    $$
        \lim_{z\rightarrow 0^+}\sum_{i=1}^r z \mathbf b_i^T (\bS_\bX + z \mathbf I)^{-1} \mathbf b_i \rightarrow c^2 \lim_{z\rightarrow 0^+} z  m(-z) = c^2 (1-1/\gamma) ~ \text{  a.s}.
    $$
    Combining the variance and the bias terms, we complete the proof.

    \subsubsection{Proof for enhanced envelope estimator}
    We use the similar techniques from the envelope estimator for both variance and bias terms. The variance term of $R(\hat\bbeta_{\bGamma}  (\lambda))$ becomes 
    $$
    \frac{\tr(\bOmega)}{n} \tr(  \bS_\bX (\bS_\bX+\lambda \mathbf I)^{-2}) \rightarrow \gamma \tr(\bOmega) \int \frac{s}{(s+\lambda)^2} F_\gamma(s). 
    $$
    
    Let $g_{n,\lambda}(\eta) = \lambda \cdot\tr(\bbeta  (\bS_\bX+\lambda(1+\eta) \mathbf I)^{-1} \bbeta^T),~\eta\in[-1/2,1/2]$. The bias term of $R(\hat\bbeta_{\bGamma}  (\lambda))$ is 
    $$
    \lambda^2 \tr(\bbeta  (\bS_\bX+\lambda \mathbf I)^{-2} \bbeta^T) = -\frac{\partial}{\partial \eta} g_n(\lambda,0).
    $$
    
    Because $$
     g_{n,\lambda}(\eta) \rightarrow \lambda c^2 m(-\lambda(1+\eta)) = \lambda c^2 \int \frac{1}{s+\lambda(1+\eta)} d F_\gamma (s),
    $$
    and derivative and limit are exchangeable, we have that 
    $$
    \lambda^2 \tr(\bbeta  (\bS_\bX+\lambda \mathbf I)^{-2} \bbeta^T) \rightarrow  \lambda^2 c^2 \int \frac{1}{(s+\lambda)^2} d F_\gamma (s).
    $$
    
    We can conclude that,
    $$
    R(\hat\bbeta_{\bGamma}  (\lambda)) \rightarrow  \int \frac{\lambda^2 c^2 + s\cdot \gamma \tr(\bOmega)}{(s+\lambda)^2} F_\gamma(s).
    $$
    The right-hand side is minimized at $\lambda^\ast = \gamma\tr(\bOmega)/c^2$. In such case, the right-hand side becomes $\gamma \tr(\bOmega)\cdot m(-\lambda^\ast)$.

\subsection{Proof of Theorem 3}
    By the distributional assumption on $\beeta$, we take expectation over $\beeta$ on $R( \hat{\bbeta}_{\bGamma}(\lambda) | \bX)$ and see that 
    \begin{equation*}
        \begin{aligned}
        \mathrm E_{\beeta} \{ R(\hat\bbeta_{\bGamma}|\bX) \}
        & = c^2/p\cdot \tr (  \Pi_\bX \bSigma_\bx \Pi_\bX )  + \frac{\tr(\bOmega)}{n} \tr( \bS_\bX^{+} \bSigma_\bx)
        \end{aligned}
    \end{equation*}
    and 
    \begin{equation*}
        \begin{aligned}
            & \mathrm E_{\beeta} \{R( \hat{\bbeta}_{\bGamma}(\lambda) | \bX) \}  = 
            \lambda^2 c^2 / p \cdot \tr(  (\bS_\bX+\lambda \mathbf I)^{-1} \bSigma_\bx (\bS_\bX+\lambda \mathbf I)^{-1} ) + \frac{\tr(\bOmega)}{n} \tr( \bSigma_\bx (\bS_\bX+\lambda \mathbf I)^{-1}\bS_\bX (\bS_\bX+\lambda \mathbf I)^{-1}).
        \end{aligned}
    \end{equation*}
    \subsubsection{Proof for enhanced envelope estimator}
    We can rewrite $\mathrm E_{\beeta} \{R( \hat{\bbeta}_{\bGamma}(\lambda) | \bX) \}$ as 
    \begin{equation*}
        \begin{aligned}
            &\mathrm E_{\beeta} \{ R( \hat{\bbeta}_{\bGamma}(\lambda) | \bX) \} = 
            (\lambda^2 c^2 / p-\lambda \tr(\bOmega)/n) \cdot \tr( \bSigma_\bx (\bS_\bX+\lambda \mathbf I)^{-2} ) + \frac{\tr(\bOmega)}{n} \tr( \bSigma_\bx (\bS_\bX+\lambda \mathbf I)^{-1}) .
        \end{aligned}
    \end{equation*}
    By the equation (4) and Lemma 2.2 from \cite{dobriban2018high}, we arrive at 
    \begin{equation*}
        \begin{aligned}
            \mathrm E_{\beeta} \{R( \hat{\bbeta}_{\bGamma}(\lambda) | \bX) \}  & \rightarrow 
            (\lambda c^2/\gamma - \tr(\bOmega)) \cdot \frac{v(-\lambda)-\lambda v'(-\lambda)}{\lambda v^2(-\lambda)} + \tr(\bOmega)\left( \frac{1}{\lambda v(-\lambda)} -1 \right)~ \text{  a.s.}.
        \end{aligned}
    \end{equation*}    
    The right-hand side can be rewritten as
    $$
     \frac{1}{\lambda v(-\lambda)} \left( \tr(\bOmega) + (\lambda c^2/\gamma - \tr(\bOmega))(1-\lambda v'(-\lambda)/v(-\lambda)) \right) - \tr(\bOmega).
    $$
    By Theorem 2.1 of \cite{dobriban2018high}, we can conclude that the above equation is minimized at $\lambda^\ast = \gamma \tr(\bOmega)/c^2$. With $\lambda^\ast$,  
    \begin{equation*}
        \begin{aligned}
            \mathrm E_{\beeta} \{ R( \hat{\bbeta}_{\bGamma}(\lambda^\ast) | \bX) \}  & \rightarrow 
            \tr(\bOmega) \left( \frac{1}{\lambda^\ast v(-\lambda^\ast)} -1 \right) ~ \text{  a.s.}.
        \end{aligned}
    \end{equation*}      
 
    \subsubsection{Proof for envelope estimator}
    From the proof of Theorem 4 of \cite{hastie2019surprises}, we see that 
    \begin{equation*}
        \begin{aligned}
        \frac{1}{p}\cdot \tr (  \Pi_\bX \bSigma_\bx \Pi_\bX )  \rightarrow \frac{1}{\gamma} \lim_{z\rightarrow 0^+} \frac{1}{v(-z)}
        \end{aligned}
    \end{equation*}      
    and 
        \begin{equation*}
        \begin{aligned}
        \frac{1}{n} \tr( \bS_\bX^{+} \bSigma_\bx) \rightarrow \lim_{z\rightarrow 0^+} \frac{v'(-z)}{v(-z)^2} -1.
        \end{aligned}
    \end{equation*} 
    Applying those two results, we prove the theorem.

\section{Additional simulation results and discussions}

    Tables B.1-B.3 are the bias and variance components of the prediction risk and Tables C.1-C.3 present the envelope dimension selected by cross-validation.
    
    Regarding bias and variance, as shown in Tables B.1-B.3, when the true $u$ is used, the bias component of the envelope estimator consistently exhibits smaller values compared to that of the enhanced envelope estimator. In contrast, the variance component of the envelope estimator is consistently larger than that of the enhanced envelope estimator. Especially, the variance component for the envelope estimator is significantly larger in cases where $p \approx n$ and $p > n$. For the enhanced envelope estimator, the bias component remains relatively stable, regardless of whether the true $u$ or a cross-validation-selected $u$ is used. However, for the envelope estimator, there is a significant increase in bias values and a decrease in variance when $u$ is chosen by cross-validation, particularly when $p$ is close to or greater than $n$. Our experiments suggest that this is due to the tendency of the envelope estimator to select a smaller envelope dimension in such scenarios to reduce the prediction risk overall by reducing the variance by sacrificing the bias of the prediction risk.
    
    Regarding the envelope dimension, as shown in Tables C.1-C.3, for the enhanced envelope estimator with cross-validation selected $u$, as the sample size increases, the percentage that the cross-validation choose the correct dimension increases. The envelope estimator with the case of $p<n$ has a similar trend. However, for the envelope estimator with cases $p\approx n$ and $p>n$, the cross-validation tends to choose a smaller envelope dimension. 

\bigskip
 
\renewcommand{\thetable}{B.1}
  
    \begin{table}[ht!]
    \scriptsize 
    \begin{center}
        \caption{\label{tab:sim:less2a} {\bf{(Cases of $\bm{p < n}$)}} The bias and variance components of the prediction risk. }
    \begin{tabular}{|c|c||cc|cc|cc|cc|} \hline
     \multicolumn{2}{|l|}{} & 
     \multicolumn{2}{|c|}{\begin{tabular}[c]{@{}c@{}}Enhanced envelope \\(using true $u$)\end{tabular}} & \multicolumn{2}{|c|}{\begin{tabular}[c]{@{}c@{}}Envelope\\ (using true $u$)\end{tabular}} &
    \multicolumn{2}{|c|}{\begin{tabular}[c]{@{}c@{}}Enhanced envelope \\(cv selected $u$)\end{tabular}} & \multicolumn{2}{|c|}{\begin{tabular}[c]{@{}c@{}}Envelope\\ (cv selected $u$)\end{tabular}} \\ \hline \hline 
    $n$ & $p$ & bias$^2$ & variance & bias$^2$ & variance & bias$^2$ & variance & bias$^2$ & variance\\ \hline \hline
    \multicolumn{10}{|l|}{{\bf Example 1.1}: ~$p/n=0.1$, $\rho=0$, $u=2$} \\ \hline 
    50&5    & 0.27 & 1.28 & 0.02 & 1.62 & 0.27 & 1.40 & 0.02 & 1.84 \\ \hline
    90&9    & 0.21 & 1.12 & 0.01 & 1.45 & 0.21 & 1.23 & 0.01 & 1.59 \\ \hline
    200&20  & 0.16 & 0.99 & 0.01 & 1.31 & 0.16 & 1.05 & 0.01 & 1.35\\ \hline \hline
    \multicolumn{10}{|l|}{{\bf Example 1.2}: ~$p/n=0.3$, $\rho=0$, $u=2$} \\ \hline 
    50&15    & 1.70 & 2.31 & 0.06 & 5.46 & 1.62 & 2.58 & 1.13 & 4.54 \\ \hline
    90&27    & 1.10 & 2.08 & 0.06 & 5.21 & 1.14 & 2.33 & 0.37 & 5.24 \\ \hline
    200&60   & 1.06 & 2.00 & 0.05 & 4.67 & 1.08 & 2.05 & 0.22 & 4.82\\ \hline \hline
    \multicolumn{10}{|l|}{{\bf Example 1.3}: ~$p/n=0.1$, $\rho=0.8$, $u=2$} \\ \hline\hline 
    50&5    &  0.55 & 0.87 & 0.02 & 1.66 & 0.49 & 1.00 & 0.14 & 1.94\\ \hline
    90&9    &  0.36 & 0.78 & 0.01 & 1.44 & 0.38 & 0.91 & 0.01 & 1.64 \\ \hline
    200&20  & 0.29 & 0.65 & 0.01 & 1.33 & 0.29 & 0.71 & 0.03 & 1.47 \\ \hline
    \multicolumn{10}{|l|}{{\bf Example 1.4}: ~$p/n=0.3$, $\rho=0.8$, $u=2$} \\ \hline\hline 
    50&15    & 0.88 & 1.39 & 0.07 & 5.47 & 0.94 & 1.59 & 1.35 & 3.23 \\ \hline
    90&27    & 1.00 & 1.27 & 0.08 & 5.28 & 1.02 & 1.51 & 1.45 & 2.96 \\ \hline
    200&60   & 0.73 & 1.17 & 0.05 & 4.70 & 0.74 & 1.21 & 0.16 & 4.95 \\ \hline
    \end{tabular}
    \end{center}
    \end{table}

\renewcommand{\thetable}{B.2}
    \begin{table}[ht!]
    \scriptsize 
    \begin{center}
        \caption{\label{tab:sim:approx2a} {\bf{(Cases of $\bm{p \approx n}$)}} The bias and variance components of the prediction risk.}
    \begin{tabular}{|c|c||cc|cc|cc|cc|} \hline
     \multicolumn{2}{|l|}{} & 
     \multicolumn{2}{|c|}{\begin{tabular}[c]{@{}c@{}}Enhanced envelope \\(using true $u$)\end{tabular}} & \multicolumn{2}{|c|}{\begin{tabular}[c]{@{}c@{}}Envelope\\ (using true $u$)\end{tabular}} &
    \multicolumn{2}{|c|}{\begin{tabular}[c]{@{}c@{}}Enhanced envelope \\(cv selected $u$)\end{tabular}} & \multicolumn{2}{|c|}{\begin{tabular}[c]{@{}c@{}}Envelope\\ (cv selected $u$)\end{tabular}} \\ \hline \hline 
    $n$ & $p$ & bias$^2$ & variance & bias$^2$ & variance & bias$^2$ & variance & bias$^2$ & variance\\ \hline \hline
    \multicolumn{10}{|l|}{{\bf Example 2.1}: ~$p/n=0.9$, $\rho=0$, $u=2$} \\ \hline 
    50&45    & 5.65 & 2.08 & 1.39 & 160.45 & 5.36 & 2.45 & 9.90 & 0.28\\ \hline
    90&80    & 4.95 & 2.03 & 1.19 & 112.33 & 4.97 & 2.16 & 10.00 & 0.00 \\ \hline
    200&180  & 4.30 & 1.89 & 1.30 & 113.93 & 4.32 & 2.02 & 10.00 & 0.00 \\ \hline \hline
    \multicolumn{10}{|l|}{{\bf Example 2.2}: ~$p/n=1.1$, $\rho=0$, $u=2$} \\ \hline 
    50&55    & 5.75 & 2.18 & 3.02 & 59.53 & 5.57 & 2.49 & 7.75 & 3.61 \\ \hline
    90&99    & 5.60 & 1.74 & 2.32 & 72.79 & 5.57 & 1.95 & 8.74 & 2.10 \\ \hline
    200&220   & 4.57 & 1.90 & 1.83 & 92.88 & 4.64 & 2.03 & 9.62 & 0.67 \\ \hline \hline
    \multicolumn{10}{|l|}{{\bf Example 2.3}: ~$p/n=0.9$, $\rho=0.8$, $u=2$} \\ \hline\hline 
    50&45    & 3.12 & 2.34 & 1.92 & 160.21 & 3.00 & 2.62 & 12.12 & 2.02\\ \hline
    90&81    & 1.86 & 2.07 & 1.21 & 112.22 & 1.86 & 2.22 & 11.59 & 0.00 \\ \hline
    200&180  & 2.37 & 1.39 & 1.26 & 115.27 & 2.40 & 1.48 & 9.95 & 0.00 \\ \hline
    \multicolumn{10}{|l|}{{\bf Example 2.4}: ~$p/n=1.1$, $\rho=0.8$, $u=2$} \\ \hline\hline 
    50&55    & 2.52 & 2.48 & 1.37 & 65.64 & 2.48 & 2.76 & 8.46 & 6.21 \\ \hline
    90&99    & 2.03 & 2.05 & 1.07 & 78.83 & 2.09 & 2.35 & 7.74 & 3.76 \\ \hline
    200&220  & 2.41 & 1.66 & 1.47 & 97.59 & 2.51 & 1.82 & 8.94 & 2.34\\ \hline
    \end{tabular}
    \end{center}
    \end{table}

\renewcommand{\thetable}{B.3} 
    \begin{table}[ht!]
    \scriptsize 
    \begin{center}
        \caption{\label{tab:sim:greater2a} {\bf{(Cases of $\bm{p > n}$)}} The bias and variance components of the prediction risk.}
    \begin{tabular}{|c|c||cc|cc|cc|cc|} \hline
     \multicolumn{2}{|l|}{} & 
     \multicolumn{2}{|c|}{\begin{tabular}[c]{@{}c@{}}Enhanced envelope \\(using true $u$)\end{tabular}} & \multicolumn{2}{|c|}{\begin{tabular}[c]{@{}c@{}}Envelope\\ (using true $u$)\end{tabular}} &
    \multicolumn{2}{|c|}{\begin{tabular}[c]{@{}c@{}}Enhanced envelope \\(cv selected $u$)\end{tabular}} & \multicolumn{2}{|c|}{\begin{tabular}[c]{@{}c@{}}Envelope\\ (cv selected $u$)\end{tabular}} \\ \hline \hline 
    $n$ & $p$ & bias$^2$ & variance & bias$^2$ & variance & bias$^2$ & variance & bias$^2$ & variance\\ \hline \hline
    \multicolumn{10}{|l|}{{\bf Example 3.1}: ~$p/n=3$, $\rho=0$, $u=2$} \\ \hline 
    50&150    & 8.24 & 1.53 & 7.65 & 5.12 & 8.22 & 1.66 & 8.21 & 2.61\\ \hline
    90&270    & 8.44 & 0.70 & 6.98 & 5.50 & 8.26 & 0.95 & 8.39 & 1.60\\ \hline
    200&600   & 7.68 & 0.96 & 6.72 & 5.19 & 7.68 & 1.05 & 7.75 & 2.16\\ \hline \hline
    \multicolumn{10}{|l|}{{\bf Example 3.2}: ~$p/n=10$, $\rho=0$, $u=2$} \\ \hline 
    50&500    & 9.29 & 0.74 & 9.19 & 1.30& 9.29 & 0.88& 9.27 &1.10 \\ \hline
    90&900    & 9.22 & 0.63 & 9.09 & 1.27& 9.20 & 0.79& 9.19 &1.17\\ \hline
    200&2000  & 9.34 & 0.38 & 9.05 & 1.49& 9.51 & 0.36& 9.68 &0.38\\ \hline \hline
    \multicolumn{10}{|l|}{{\bf Example 3.3}: ~$p/n=3$, $\rho=0.8$, $u=2$} \\ \hline\hline 
    50&150    & 6.50 & 1.34 & 3.56 & 11.7 & 6.59 & 1.37 & 8.10 & 1.82\\ \hline
    90&270    & 4.98 & 1.74 & 2.79 & 11.77 & 4.87 & 1.95 & 4.94 & 3.28 \\ \hline
    200&600   & 4.42 & 1.61 & 2.64 & 11.25 & 4.44 & 1.71 & 5.56 & 2.46\\ \hline
    \multicolumn{10}{|l|}{{\bf Example 3.4}: ~$p/n=10$, $\rho=0.8$, $u=2$} \\ \hline\hline 
    50&500    & 7.60 & 1.69 & 7.03 & 4.62 & 7.50 & 1.63 & 7.31 & 2.47 \\ \hline
    90&900    & 7.85 & 1.31 & 7.03 & 4.80 & 7.97 & 1.36 & 8.02 & 1.88 \\ \hline
    200&2000  & 7.58 & 1.04 & 6.57 & 4.55 & 7.67 & 1.07 & 7.98 & 2.49 \\ \hline
    \end{tabular}
    \end{center}
    \end{table}

\renewcommand{\thetable}{C.1}
  
    \begin{table}[ht!]
    \scriptsize 
    \begin{center}
        \caption{\label{tab:sim:less2b} {\bf{(Cases of $\bm{p < n}$)}} The distribution of $\hat u$, where $\hat u$ is selected by cross-validation and $\#(\hat u=j)=\#(j)$ is the number of times that cross-validation selects envelope dimension $j$.}
    \begin{tabular}{|c|c||cccc||cccc|} \hline
     \multicolumn{2}{|l|}{} & 
    \multicolumn{4}{|c|}{\begin{tabular}[c]{@{}c@{}}Enhanced envelope \\(cv selected $u$)\end{tabular}} & \multicolumn{4}{|c|}{\begin{tabular}[c]{@{}c@{}}Envelope\\ (cv selected $u$)\end{tabular}} \\ \hline \hline 
    $n$ & $p$ & $\#(0)$ & $\#(1)$ & $\#(2)$ & $\#(3)$ & $\#(0)$ & $\#(1)$ & $\#(2)$ & $\#(3)$ \\ \hline \hline
    \multicolumn{10}{|l|}{{\bf Example 1.1}: ~$p/n=0.1$, $\rho=0$, $u=2$} \\ \hline 
    50&5    & 0 & 1 & 78 & 21 & 0 & 2 & 85 & 13 \\ \hline
    90&9    & 0 & 0 & 91 & 9 & 0 & 0& 93 & 7 \\ \hline
    200&20  & 0 & 0 & 95 & 5 & 0 & 0 & 98 & 2 \\ \hline \hline
    \multicolumn{10}{|l|}{{\bf Example 1.2}: ~$p/n=0.3$, $\rho=0$, $u=2$} \\ \hline 
    50&15    & 0 & 20 & 63 & 17 & 0 & 55 & 42 & 3 \\ \hline
    90&27    & 0 & 2 & 83 & 15 & 0 & 29 & 69 & 2 \\ \hline
    200&60   & 0 & 0 & 96 & 4 & 0 & 20 & 80 & 0 \\ \hline \hline
    \multicolumn{10}{|l|}{{\bf Example 1.3}: ~$p/n=0.1$, $\rho=0.8$, $u=2$} \\ \hline\hline 
    50&5    &  1 & 15 & 60 & 24 & 11 & 25 & 52 & 12\\ \hline
    90&9    &  0 & 2 & 75 & 23 & 0 & 6 & 88 & 6 \\ \hline
    200&20  & 0 & 0 & 91 & 9 & 0 & 7 & 90 & 3  \\ \hline
    \multicolumn{10}{|l|}{{\bf Example 1.4}: ~$p/n=0.3$, $\rho=0.8$, $u=2$} \\ \hline\hline 
    50&15    & 0 & 18 & 60 & 22 & 0 & 73 & 27 & 0 \\ \hline
    90&27    & 0 & 10 & 75 & 15 & 0 & 74 & 26 & 0 \\ \hline
    200&60   & 0 & 0 & 95 & 5 & 0 & 16 & 84 & 0 \\ \hline
    \end{tabular}
    \end{center}
    \end{table}

\renewcommand{\thetable}{C.2}
    \begin{table}[ht!]
    \scriptsize 
    \begin{center}
        \caption{\label{tab:sim:approx2b} {\bf{(Cases of $\bm{p \approx n}$)}} The distribution of $\hat u$, where $\hat u$ is selected by cross-validation and $\#(\hat u=j)=\#(j)$ is the number of times that cross-validation selects envelope dimension $j$.}
    \begin{tabular}{|c|c||cccc||cccc|} \hline
     \multicolumn{2}{|l|}{} & 
    \multicolumn{4}{|c|}{\begin{tabular}[c]{@{}c@{}}Enhanced envelope \\(cv selected $u$)\end{tabular}} & \multicolumn{4}{|c|}{\begin{tabular}[c]{@{}c@{}}Envelope\\ (cv selected $u$)\end{tabular}} \\ \hline \hline 
        $n$ & $p$ & $\#(0)$ & $\#(1)$ & $\#(2)$ & $\#(3)$ & $\#(0)$ & $\#(1)$ & $\#(2)$ & $\#(3)$ \\ \hline \hline
    \multicolumn{10}{|l|}{{\bf Example 2.1}: ~$p/n=0.9$, $\rho=0$, $u=2$} \\ \hline 
    50&45    & 0 & 35 & 43 & 22 & 98 & 2 & 0 & 0  \\ \hline
    90&80    & 0 & 12 & 74 & 14 & 100 & 0 & 0 & 0  \\ \hline
    200&180  & 0 & 12 & 82 & 6 & 100 & 0 & 0 & 0  \\ \hline \hline
    \multicolumn{10}{|l|}{{\bf Example 2.2}: ~$p/n=1.1$, $\rho=0$, $u=2$} \\ \hline 
    50&55    & 0 & 32 & 51 & 17 & 64 & 35 & 1 & 0  \\ \hline
    90&99    & 0 & 35 & 56 & 9 & 82 & 18 & 0 & 0  \\ \hline
    200&220  & 0 & 19 & 76 & 5 & 96 & 4 & 0 & 0\\ \hline \hline
    \multicolumn{10}{|l|}{{\bf Example 2.3}: ~$p/n=0.9$, $\rho=0.8$, $u=2$} \\ \hline\hline 
    50&45    & 0 & 23 & 56 & 21 & 92 & 8 & 0 & 0 \\ \hline
    90&81    & 0 & 1 & 86 & 13 & 100 & 0 & 0 & 0  \\ \hline
    200&180  & 0 & 5 & 89 & 6 & 100 & 0 & 0 & 0 \\ \hline
    \multicolumn{10}{|l|}{{\bf Example 2.4}: ~$p/n=1.1$, $\rho=0.8$, $u=2$} \\ \hline\hline 
    50&55    & 0 & 4 & 72 & 24 & 60 & 36 & 4 & 0 \\ \hline
    90&99    & 0 & 10 & 76 & 14 & 73 & 27 & 0 & 0 \\ \hline
    200&220  & 0 & 21 & 74 & 5 & 88 & 12 & 0 & 0 \\ \hline
    \end{tabular}
    \end{center}
    \end{table}
    
\renewcommand{\thetable}{C.3}
    \begin{table}[ht!]
    \scriptsize 
    \begin{center}
        \caption{\label{tab:sim:greater2b} {\bf{(Cases of $\bm{p > n}$)}} The distribution of $\hat u$, where $\hat u$ is selected by cross-validation and $\#(\hat u=j)=\#(j)$ is the number of times that cross-validation selects envelope dimension $j$.}
    \begin{tabular}{|c|c||cccc||cccc|} \hline
     \multicolumn{2}{|l|}{} & 
    \multicolumn{4}{|c|}{\begin{tabular}[c]{@{}c@{}}Enhanced envelope \\(cv selected $u$)\end{tabular}} & \multicolumn{4}{|c|}{\begin{tabular}[c]{@{}c@{}}Envelope\\ (cv selected $u$)\end{tabular}} \\ \hline \hline 
        $n$ & $p$ & $\#(0)$ & $\#(1)$ & $\#(2)$ & $\#(3)$ & $\#(0)$ & $\#(1)$ & $\#(2)$ & $\#(3)$ \\ \hline \hline
    \multicolumn{10}{|l|}{{\bf Example 3.1}: ~$p/n=3$, $\rho=0$, $u=2$} \\ \hline 
    50&150    & 1 & 43 & 34 & 22 & 17 & 51 & 28 & 4 \\ \hline
    90&270    & 1 & 41 & 34 & 24 & 34 & 55 & 10 & 1 \\ \hline
    200&600   & 0 & 20 & 67 & 13 & 19 & 61 & 20 & 0\\ \hline \hline
    \multicolumn{10}{|l|}{{\bf Example 3.2}: ~$p/n=10$, $\rho=0$, $u=2$} \\ \hline 
    50&500    & 5 & 31 & 33 & 31 & 11 & 34 & 32 & 23 \\ \hline
    90&900    & 4 & 14 & 41 & 41 & 19 & 25 & 32 & 24 \\ \hline
    200&2000  & 31 & 30 & 17 & 22 & 69 & 22 & 7 & 2\\ \hline \hline
    \multicolumn{10}{|l|}{{\bf Example 3.3}: ~$p/n=3$, $\rho=0.8$, $u=2$} \\ \hline\hline 
    50&150    & 9 & 27 &  35 & 29 & 78 & 15 & 6 & 1 \\ \hline
    90&270    & 0 & 36 & 51 & 13 & 4 & 89 & 7 & 0 \\ \hline
    200&600   & 0 & 21 & 75 & 4 & 3 & 96 & 1 & 0\\ \hline
    \multicolumn{10}{|l|}{{\bf Example 3.4}: ~$p/n=10$, $\rho=0.8$, $u=2$} \\ \hline\hline 
    50&500    & 1 & 46 & 35 & 18 & 9 & 63 & 23 & 5  \\ \hline
    90&900    & 2 & 54 & 32 & 12 & 10 & 74 & 11 & 5  \\ \hline
    200&2000  & 1 & 10 & 85 & 4  & 44 & 24 & 32 & 0 \\ \hline
    \end{tabular}
    \end{center}
    \end{table}

$~$
\newpage 
$~$
\newpage
$~$
\newpage
$~$
\newpage
$~$
\newpage 
$~$
\newpage

\end{appendices}

\clearpage

\bibliographystyle{asa.bst} 
\bibliography{reference}

\end{document}